\documentclass[a4paper,11pt]{article}
\pdfoutput=1
\usepackage{jcappub}
\usepackage{setspace}  

\usepackage{tikz,xcolor,hyperref}
\usepackage[normalem]{ulem}
\usepackage{tikzsymbols}
\usepackage{float}
\usepackage{bm}
\usepackage{xcolor} 
\usepackage{graphicx}
\usepackage[utf8]{inputenc}
\usepackage[T1]{fontenc}
\usepackage{array}
\usepackage{graphicx,rotate,multicol,caption}
\usepackage{xcolor,colortbl}
\usepackage{multirow,tabularx}
\usepackage{longtable}
\usepackage{lscape}
\usepackage{booktabs}
\usepackage{float}
\usepackage{hyperref}
\usepackage{pifont}
\usepackage{caption}
\DeclareCaptionLabelSeparator{none}{ - }
\usepackage{parskip}
\usepackage{geometry} 
\usepackage{amsmath}
\usepackage{verbatim}

\definecolor{Celadon}{rgb}{0.67, 0.94, 0.82}

\newcommand{\omi}[1]{}

\newcounter{mysubequation}[equation]

\usepackage[bitstream-charter]{mathdesign}
\usepackage[nointegrals]{wasysym} 

\renewcommand{\figurename}{FIGURE}

\usepackage{lmodern} 
\usepackage{amsmath}                                                    
\usepackage{amsthm}                                                     
\usepackage{multirow}
\usepackage[nameinlink]{cleveref}
\Crefname{figure}{Fig.}{Figs.}

\usepackage{adjustbox}
\usepackage{color, colortbl} 
\definecolor{Gray}{gray}{0.9}
\usepackage{ragged2e}
\usepackage{titlesec}
\usepackage{cellspace}
\usepackage{makecell}
\usepackage{setspace}
\usepackage[nameinlink]{cleveref}
\Crefname{figure}{Fig.}{Figs.}

\newcommand{\ag}[1]{{\color{orange}[Anish: #1]}}
\usepackage{subfig}
\usepackage{graphicx}
\usepackage{mathtools}
\usepackage{orcidlink}
\usepackage{natbib}
\def\beq{\beq\begin{align}}
\def\eeq{\end{align}\eeq}

\newcommand{\Cos}{\text{Cos}}

\def\beq{\begin{equation}\begin{align}}
\def\eeq{\end{align}\end{equation}}

\begin{document}
\title{Primordial Black Holes and Second-order Gravitational Waves in Axion-like Hybrid Inflation}
\author[]{Waqas Ahmed \orcidlink{0000-0001-8136-9958}$^{a\dagger}$,}
\author[]{Anish Ghoshal \orcidlink{0000-0001-7045-302X}$^{b^{\heartsuit}}$}
\author[]{Umer Zubair \orcidlink{0000-0003-1671-8722}$^{c^{*}}$}

\affiliation[a]{Center for Fundamental Physics and School of Mathematics and Physics, Hubei Polytechnic University, Huangshi 435003, China}
\affiliation[b]{ Institute of Theoretical Physics, Faculty of Physics,\\ University of Warsaw,
ul. Pasteura 5, 02-093 Warsaw, Poland}
\affiliation[c]{Department of Physics and Astronomy, University of Delaware, Newark, DE 19716, USA}

\emailAdd{waqasmit@hbpu.edu.cn.$^\dagger$}
\emailAdd{anish.ghoshal@fuw.edu.pl$^{\heartsuit}$}
\emailAdd{umer@udel.edu$^{*}$}

\abstract {We investigate the possibility that primordial black holes (PBHs) can be formed from large curvature perturbations  generated during the waterfall phase transition in a hybrid inflation model driven by an axion-like particle (ALP) $\phi$. The model predicts a spectral index $n_s \simeq 0.964$ and a tensor-to-scalar ratio $r \simeq 0.003$, in agreement with Planck data and potentially testable by next generation cosmic microwave background (CMB) experiments such as CMB-S4 and LiteBIRD. We find that the PBH mass and the peak of the associated scalar-induced gravitational wave (SIGW) spectrum are correlated with the ALP mass. In particular, PBHs in the mass range $10^{-13}\, M_\odot$ can constitute either the entire dark matter (DM) content of the universe or a significant fraction of it. The predicted second-order GWs from this mechanism are within the sensitivity reach of future observatories like LISA and ET. The typical reheating temperature in the model is around $10^6 - 10^7$ GeV is consistent with Big Bang Nucleosynthesis (BBN) constraints.}

 
\maketitle

\section{Introduction}
\label{sec:intro}



Two scales that one may infer from cosmology are the temperature fluctuations seen in the Cosmic Microwave Background (CMB) scale ($k = 0.056 MPc^{-1}$) also known as the pivot scale and the scale of during cosmic inflation known as Hubble scale. Besides, recently there has also been a lots of interest at scales much smaller than that of the pivot scale, which correspond to the late stages of the cosmic inflationary paradigm. At these small scales of the universe there may exist large peaks in the amplitude of scalar perturbations which could lead to the formation of primordial black holes (PBHs), which may explain the observations of the supermassive ~\cite{LyndenBell:1969yx, Kormendy:1995er} and stellar-mass black hole (BH) merger events detected due to the observation of gravitational wave (GW) in LIGO-VIRGO-KAGRA~\cite{LIGOScientific:2016dsl,LIGOScientific:2021djp}. At the same time, associated with these large peaks in the amplitude of the curvature perturbation, is the production of a stochastic gravitational wave background (SGWB) at the second-order in perturbation theory which also known as scalar-induced gravitational waves (SIGW)~\cite{Hawking:1971ei,Carr:1974nx,Carr:1975qj}. These large density fluctuations are quite common in inflationary scenarios, for instance, when the inflaton undergoes a period of ultra-slow-roll while tracersing a very flat region of its potential (see ref.~\cite{Escriva:2022duf} for a recent review on PBHs) but sometime may involve some higher degrees of fine-tuning see Ref. \cite{Cole:2023wyx} for details. One of the alternative to ultra-slow roll scenarios that could also produce exceptionally large density fluctuations, is hybrid inflationary paradigm \cite{Linde:1991km,Linde:1993cn} where it can be achieved, see Ref.~\cite{Garcia-Bellido:1996mdl}. The level of fine-tuning is milder there-in \cite{Afzal:2024hwj,Afzal:2024xci,Spanos:2021hpk}.


 In context to the Standard Model (SM) of particle physics, something well-known as the Strong CP problem is one of the unsolved problems; and it strongly recommends to go beyond the SM (BSM). The so-called Peccei-Quinn (PQ) mechanism~\cite{Peccei:1977hh,Peccei:1977ur}, which cosniders the existence of a light pseudo-Nambu-Goldstone boson (pNGB), famously called the QCD axion, dynamically makes this strong CP phase $\bar{\theta}_{\rm QCD}$ to be zero or vanishingly small~\cite{Weinberg:1977ma,Wilczek:1977pj}. 
At the same time, due to the non-perturbative effects of QCD, strong confining dynamics generates QCD axion mass which must be lighter than $O(10)$ meV to satisfy the current astrophysical observational 
bounds~\cite{Hamaguchi:2018oqw,Beznogov:2018fda,Leinson:2019cqv}. (see Refs.\,\cite{Jaeckel:2010ni,Ringwald:2012hr,Arias:2012az,Graham:2015ouw,Marsh:2015xka,Irastorza:2018dyq, DiLuzio:2020wdo} for reviews).

If one goes beyond the standard QCD axion paradigm, there exist several Axion-like
particles (ALPs) that are also light gauge-singlet pseudoscalar bosons which couple weakly to the
Standard Model (SM) and generically appear as the pseudo-Nambu-Goldstone boson (pNGB), particularly
in theories with a spontaneously broken global $U(1)$ symmetry, and are motivated from various string
theory constructions \cite{Arvanitaki:2009fg}. It has also been proposed that some of these ALPs may also solve some other existing puzzles in the
SM, such as the gauge hierarchy problem of the Standard Model via what is known as the relaxion
cosmology~\cite{Graham:2015cka}, ALPs may source the primordial density perturbation due to a period of cosmic inflation driven by it~\cite{Freese:1990rb,Adams:1992bn,Daido:2017wwb}, be a component of non-thermal dark matter (DM) candidates produced via vacuum misalignment mechanism~\cite{Preskill:1982cy,Abbott:1982af,Dine:1982ah}. ALPs can also explain the dark energy in the universe via quasi-flat potentials and its dynamics~\cite{Jain:2004gi,Kim:2009cp,Kim:2013jka,Lloyd-Stubbs:2018ouj}, and source the observed baryon asymmetry of the universe either directly or via
leptogenesis due ot its coupling to heavy neutrinos~\cite{Daido:2015gqa,DeSimone:2016bok}. Recently ALPs have been considered in context to
non-standard cosmology with axion-driven kination (stiffer equation of state than radiation-domination $w=1/3$ )
\cite{Co:2019jts,Co:2019wyp,Co:2020jtv,Harigaya:2021txz,Co:2021qgl}. Recently Gravitational Waves and Primordial Black Holes formation has proved to be an useful tool to probe early universe axion cosmology, from strog first-order phase transitions in Refs. \cite{DelleRose:2019pgi,Ghoshal:2020vud,Conaci:2024tlc} or from cosmological perturbation arising due to axion fluctuations in early universe \cite{Chen:2024pge} having detectable primordial non-gaussianities which can be tested in next generation of cosmological surveys \cite{Ghoshal:2023lly,Fong:2023egk,Ghoshal:2024wom}.

All couplings of ALPs to the SM particles are parametrically suppressed by $1/f$ where f represents the $U(1)_{\rm PQ}$ symmetry breaking energy scale, also known as the ALP decay constant in the literature. This energy scale is generated when the the radial part of the SM-singlet complex scalar $\Phi$ charged under $U(1)_{\rm PQ}$ acquires a vacuum expectation value (VEV). Denoting the vev by $\langle \Phi\rangle=f/\sqrt{2}$, which quite larger the electroweak scale $v_{\rm ew}\simeq 246.2$ GeV and easily avoids current observational constraints~\cite{Jaeckel:2010ni, Irastorza:2018dyq}. The $\Phi$-field can then be expressed as:
\begin{align}
\Phi(x) \ = \ \frac{1}{\sqrt 2}\left[f +\sigma(x)\right]e^{\phi(x)/f} \, .
\label{eq:phix}
\end{align}
Modulus $\phi$ of the $\Phi$-field receives a large mass term $m_\phi\sim f$. However the angular part $a$ is light and becomes the pseudo-Nambu-goldstone (pNGB) with mass $m_a$ which as usual is generated due to an explicit low energy $U(1)$-breaking non-perturbative effects. Therefore, as is the norm with in any effective-field-theory (EFT), for the low-energy phenomenology of ALPs, the heavier modulus part $\phi$ is integrated out of the theory, and thus we are left with only independent parameters being the mass of the ALP $m$ and the decay constant $f$. For the rest of the paper we will consider these as microscopic BSM independent parameters and try to identify the phenomenological consequences such as PBH and GW in relation to these quantities.

In this paper, we consider the possibility that the inflaton sector is driven by axion or axion-like particles (ALP) and participates in waterfall transition.  We demonstrate that the mass of PBHs and the peak in the gravitational wave (GW) spectrum exhibit a correlation with the mass of the axion-like particle in such scenarios.
The model predicts a spectral index of $n_s \approx 0.964$ and a tensor-to-scalar ratio of $r \approx 0.003$, both in agreement with CMB observations. We further explore that in the model, PBHs in the mass range of $10^{-13} M_\odot$ potentially constituting a significant fraction, or even the entirety, of the universe's dark matter content. Additionally, we discuss how the reheating temperature is consistent with the constraints imposed by Big Bang Nucleosynthesis (BBN), providing a coherent cosmological scenario for PBH formation and dark matter production in simple extension of the Standard Model (SM) of particle physics.

The paper is organized as follows: In Section \ref{sec2}, we review the $\alpha-$-Attractor Axionic Hybrid Inflation model, providing an overview of its key features and examining its cosmological dynamics. Section \ref{sec3} focuses on scalar perturbations, analyzing their role in the formation of primordial black holes (PBHs) and the mechanisms through which these perturbations manifest. In Section \ref{sec4}, we discuss the generation of second-order gravitational waves induced by these scalar perturbations, exploring their potential observational signatures and significance. Section \ref{sec5} provides estimates for reheating, detailing the implications for the post-inflationary universe. Finally, Section \ref{sec6} concludes with a summary of the main findings, highlighting their relevance to current cosmological theories and future research directions.

\medskip
\section{\texorpdfstring{$\alpha$-Attractor Axionic Hybrid Inflation: Exploring Cosmological Dynamics}{}}\label{sec2}
\label{alphatt}

We explore the hybrid inflation employing backgrounds of axions (or axion-like particles (ALP) in general) where the scalar field $\phi$ serves as the inflaton and $\psi$
play the role of waterfall field. Moreover, we modify the kinetic energy component of the inflaton field, denoted as $\phi$. This modification is considered within the framework of an $\alpha$-attractor model, leading to the following Lagrangian density density, \cite{Braglia:2022phb,Afzal:2024xci,Afzal:2024hwj}
\begin{align}
\label{Lagrang}
    \mathscr{L}\simeq &\dfrac{\left(\partial^\mu \phi\right)^2}{2\left(1-\dfrac{\phi^2}{6\,\alpha}\right)^2}+\dfrac{\left(\partial^\mu\psi\right)^2}{2}-V(\phi,\psi).
\end{align}
Where the potential $V(\phi,\psi)$ is define as:
\begin{equation}\label{hpoten}
     V(\psi,\phi)=\kappa^2\left(M^2-\dfrac{\psi^2}{4}\right)^2+V(\phi)+\dfrac{\lambda^2}{4}\,\phi^2\,\psi^2+b^3\, \psi ,
\end{equation} 
where
\begin{equation}
     V(\phi)=f^2 m^2\ \left[1-\Cos\left(\frac{\phi}{f}\right)\right].
\end{equation}
Here, $M$, $m$ represent the mass parameters associated with the inflaton and waterfall fields, respectively.  {\color{black} The inclusion of the linear term $b^3 \psi$ plays a crucial role in lifting the degeneracy of the $\mathbb{Z}_2$-symmetric vacua located at $\psi = \pm 2M$. This explicitly breaks the symmetry, leading to a unique global minimum for the potential and thereby preventing the formation of domain walls after symmetry breaking.
Moreover, the presence of this term ensures that the waterfall transition is smooth rather than abrupt. Near the critical point, the inflaton field $\phi$ evolves slowly, and the effective mass of $\psi$ gradually shifts from positive to tachyonic. As a result, the system undergoes a second-order-like transition instead of a sudden quench. This slow evolution suppresses the rapid growth of $\psi$ fluctuations, which in turn prevents the overproduction of primordial black holes, a problem that typically arises in models with sharp waterfall transitions. The uplifted vacuum from the $b^3 \psi$ term also slightly contributes to the post-inflationary vacuum energy, but it remains negligible and does not affect the inflationary predictions.
}  The parameter $f$  is the decay constant of the axion, which also plays the role of the inflaton in this model. Additionally, $\kappa$ and $\lambda$ are dimensionless couplings. Along the valley, the waterfall field stabilizes at $\psi=0$ as long as $\phi$ is larger than the critical field value $\phi_c$ inflaton end and the field $\psi$ falls into one of the two minima of the potential at $\psi \simeq \pm M$ depending on the sign of the coefficient of the linear term $b$. \textcolor{black}{Also the parameter $\alpha$ originates from the hyperbolic geometry of the inflaton’s field space in $\alpha$-attractor models \cite{Kallosh:2013yoa,Galante:2014ifa}. It sets the curvature scale via $R_{\text{field}} \sim -1/\alpha$. The associated kinetic term has a pole at $\phi = \sqrt{6\alpha}$, which stretches the potential in the canonical frame, modifying inflationary dynamics. Similar kinetic structures can also emerge from renormalization group effects in models with a dark Higgs or the Standard Model Higgs sector \cite{Ghoshal:2023pcx}. The parameter $\alpha$ also controls inflationary observables like $n_s$ and $r$, and its value can be constrained by recent CMB data from Planck and ACT will be explored in detail in the subsequent sections.}. The hybrid potential \cref{hpoten}, 
in terms of the canonically normalized inflaton field \cite{Kallosh:2022ggf} $\phi\rightarrow \sqrt{6\,\alpha} \text{Tanh}\left(\varphi/\sqrt{6\,\alpha}\right)$, can now be written as,
\begin{align}
\label{canonpoten}
V(\psi,\varphi) &= \kappa^2\left(M^2-\dfrac{\psi^2}{4}\right)^2+m^2f^2\left[1-\Cos\left(\frac{\sqrt{6\,\alpha}\, \text{Tanh}\left(\dfrac{\varphi}{\sqrt{6\,\alpha}}\right)}{f}\right)\right] \\\notag
&+\dfrac{\lambda^2}{2}\psi^2\left(\sqrt{6\,\alpha}\, \text{Tanh}\left(\dfrac{\varphi}{\sqrt{6\,\alpha}}\right)\right)^2 +b^3\, \psi.
\end{align}
 The mass squared of the waterfall field at $\psi=0$ is,
\begin{align}
\label{waterfallmass}
M_\psi^2=\left(-\kappa^2\,M^2+\dfrac{1}{2}\left(\lambda\,\sqrt{6\,\alpha}\,\text{Tanh}\left(\dfrac{\varphi}{\sqrt{6\,\alpha}}\right)\right)^2\right).
\end{align}
\begin{figure}[t]
    \centering
    \includegraphics[width=0.48\linewidth,height=6.7cm]{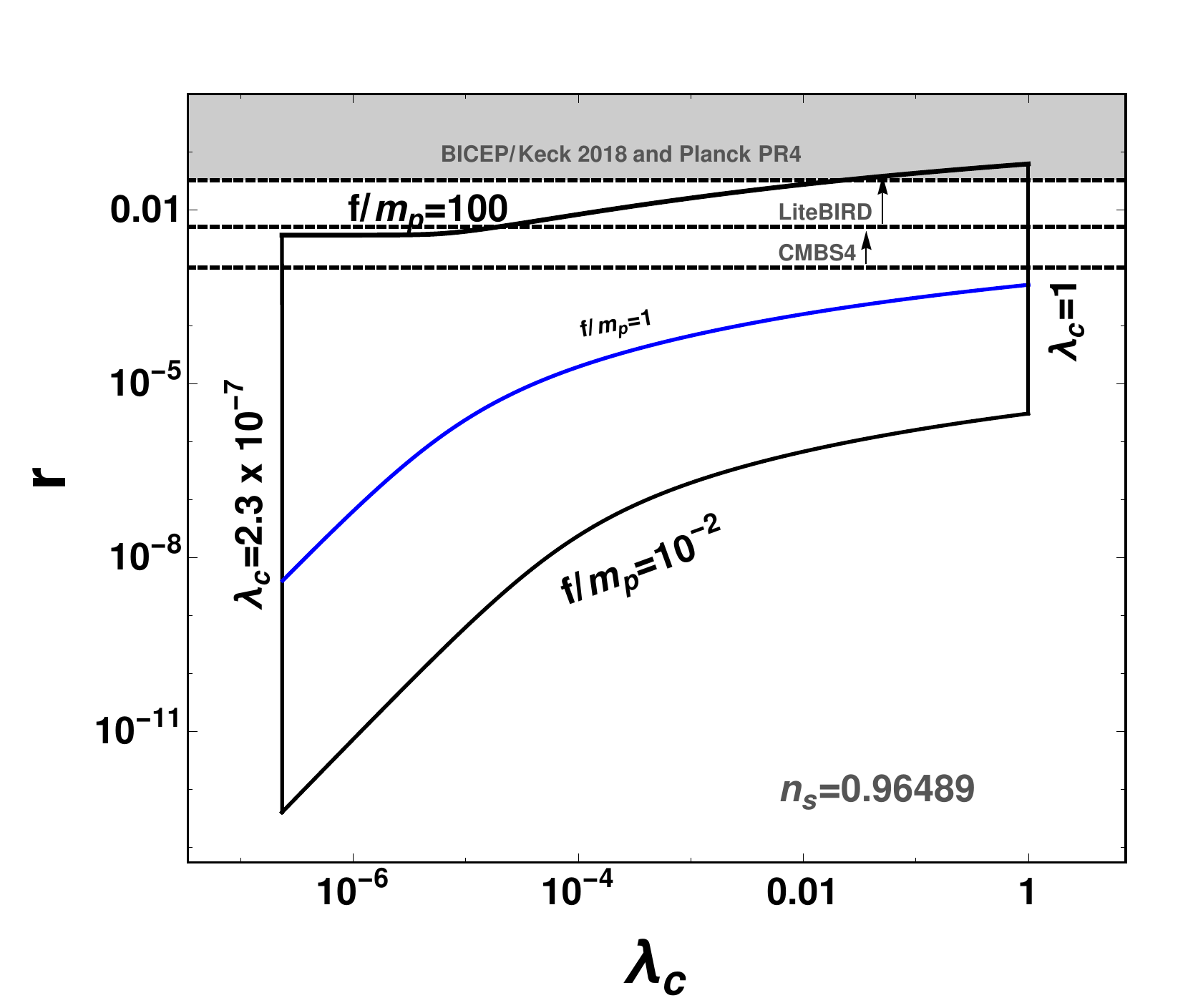}
    \quad
    \includegraphics[width=0.48\linewidth,height=6.7cm]{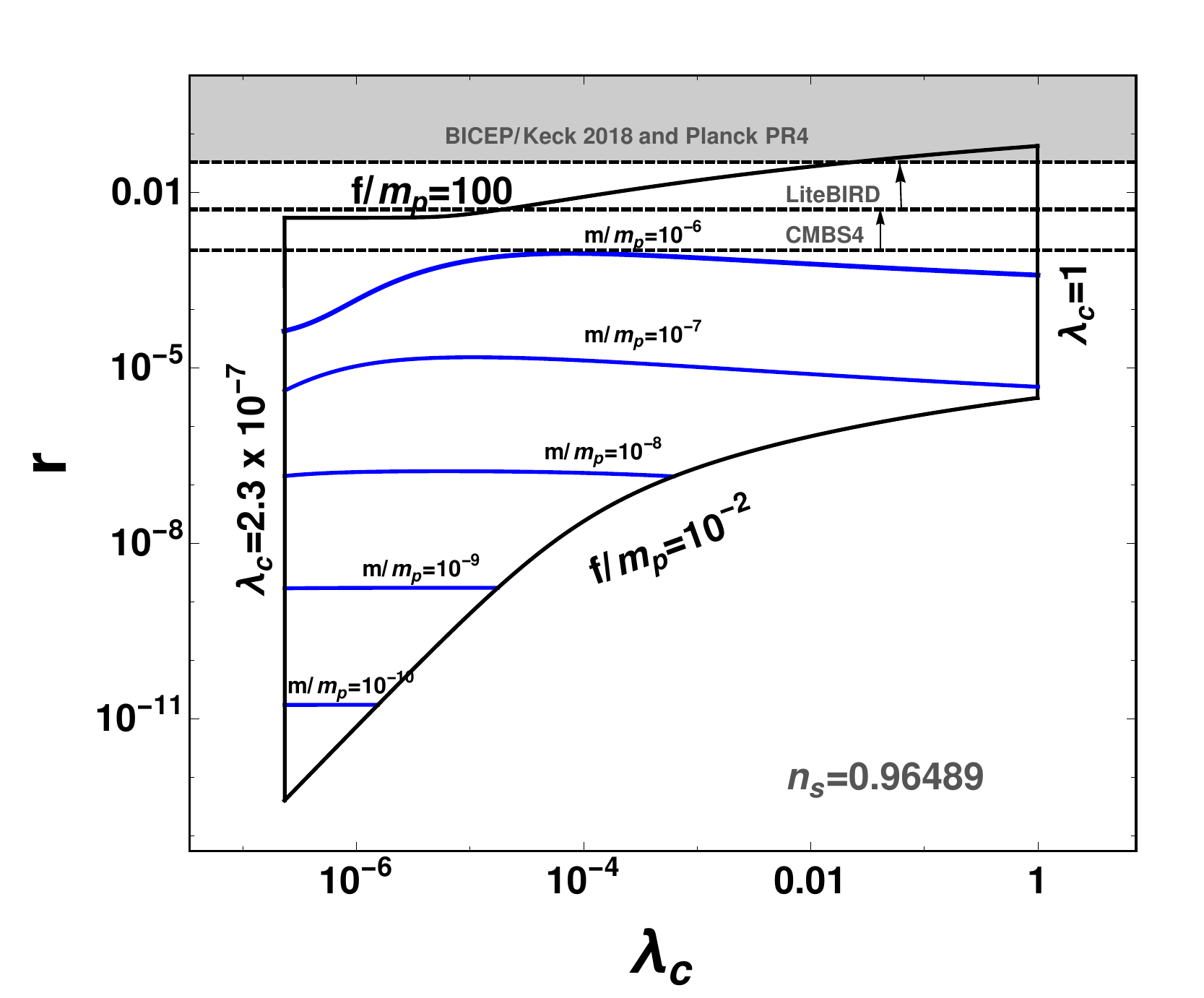}
      \quad
       \includegraphics[width=0.48\linewidth,height=6.7cm]{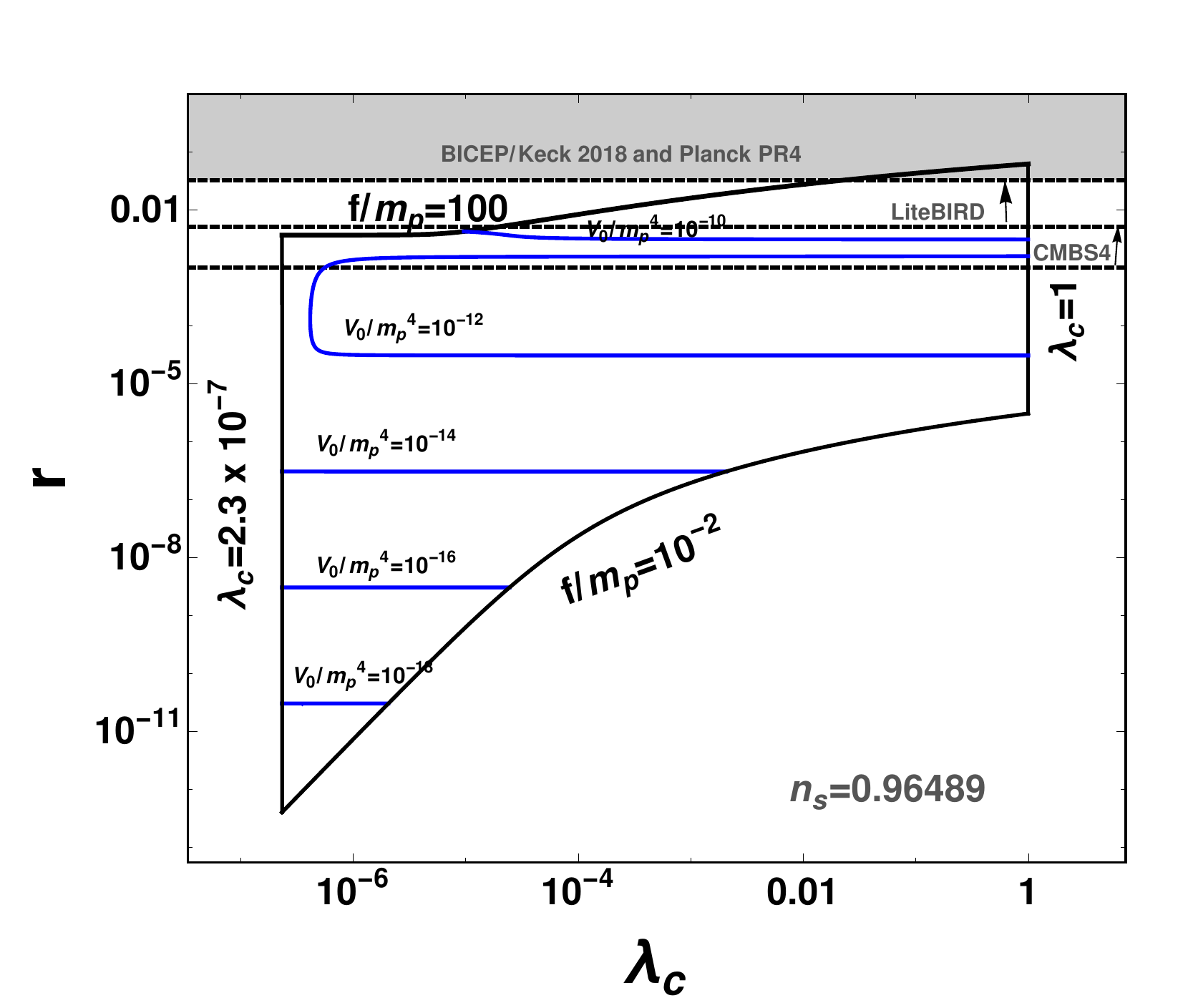}
      \quad
    \caption{\it  The relationship between the dimensionless parameter $\lambda_c$ and tensor-to-scalar ratio $r$ is depicted. The inside mesh shows the variation of $V_{0}$, mass of axions $m$, and $f$ which is the decay constant. Gray color bands illustrate constraints from the BICEP/Keck 2018 and Planck PR4 satellite \cite{Planck:2018vyg}. Projected limits from future experiments such as LiteBIRD \cite{LiteBIRD:2022cnt} and CMB-S4\cite{CMB-S4:2016ple} are also indicated.}
    \label{results1}
\end{figure}
In this paper, we will assume that $(\lambda\,\sqrt{6\,\alpha})^2/2 > \kappa^2\,M^2$ such that $M_\psi^2>0$ at large $\varphi>\varphi_c$ to stabilize the inflationary trajectory at $\psi = 0$, where,
\begin{align}
    \text{Tanh}^2\left(\dfrac{\varphi_c}{\sqrt{6\,\alpha}}\right)=\dfrac{V_{0}^{1/2}}{3\,\alpha\,\lambda_c}.
\end{align}
where $V_{0}$ is define as $V_{0}=\kappa^2\,M^4$ and $\lambda_c$ is $\lambda_c=\lambda^2/\kappa$. During inflation, as long as $\varphi\lesssim\varphi_c$, the effective mass square of $\psi$ becomes negative which gives rise to tachyonic instability that will grow the curvature perturbations. The effective single field
potential along $\psi=0$ direction given by
\begin{align}
\label{canonpoten1}
V(\varphi) &= V_{0}+m^2f^2\left[1-\cos\left(\frac{\sqrt{6\,\alpha}\, \text{Tanh}\left(\dfrac{\varphi}{\sqrt{6\,\alpha}}\right)}{f}\right)\right]. 
\end{align}
The slow-roll parameters are given by \cite{Chatterjee:2017hru},
\begin{align}
    \epsilon_V &= \dfrac{1}{2}\left(\dfrac{\partial_\varphi V}{V}\right)^2,\,\,\,\,\,\eta_V = \left(\dfrac{\partial_\varphi^2 V}{V}\right), \\\notag
    \delta_V^2 & = \left(\dfrac{\partial_\varphi V\,\partial_\varphi^3 V}{V^2}\right),\,\,\,\,\, \sigma_V^3=  \left(\dfrac{(\partial_\varphi V)^2\,\partial_\varphi^4 V}{V^3}\right).
\end{align}
Here, we used  $m_\text{Pl}\simeq 1$ unit for  reduced Planck mass. The amplitude of the scalar power spectrum is given by,
\begin{align}
A_{s}(k_0) = \frac{1}{24\,\pi^2}
\left( \frac{V(\varphi_0)}{\epsilon(\varphi_0)}\right),  \label{curv}
\end{align}
which at the pivot scale $k_0 = 0.05\, \rm{Mpc}^{-1}$ is given by $A_{s}(k_0) = 2.137 \times 10^{-9}$, as measured by Planck 2018 \cite{Planck:2018vyg}.
The last number of e-folds, $N_0$, is given by,
\begin{align}\label{Ngen}
N_0 = \int_{\varphi_c}^{\varphi_{0}}\left( \frac{V}{%
	V'}\right) dx,
\end{align}
where $x_0 \equiv x(k_0)$ and
$\varphi_c$ are the field values at the pivot scale $k_0$ and at the end of inflation, respectively.  In the slow roll limit, the spectral index $n_s$, its running and running of the running, are given by \cite{Chatterjee:2017hru},
\begin{align}
    n_s &= 1-6\,\epsilon_V + 2\,\eta_V,\,\,\,\,\, \text{d}n_s/\text{d}\ln k=16\,\epsilon_V\,\eta_V-24\,\epsilon_V^2-2\,\delta_V^2,\\\notag
    \text{d}^2n_s/\text{d}\ln k^2 &=-192\,\epsilon_V^3+192\,\epsilon_V^2\,\eta_V-32\,\epsilon_V\,\eta_V^2-24\,\epsilon_V\delta_V^2+2\,\eta_V\delta_V^2+2\sigma_V^3.
\end{align}
In order to discuss the predictions of the model, some discussion of the effective number of independent parameters is in order. Apart from the $m$ and $f$ parameters of the axion potential, the fundamental parameters of the potential in Eq \eqref{canonpoten1} are $\kappa$ and $M$, , which can be reduced to $V_{0}$ and $\phi_{c}$ for the effective potential in Eq. \eqref{canonpoten1}. We, however, take $V_{0}$, and $\kappa_c$ as the
effective independent parameters with $\varphi_c$ as.
\begin{align}
\label{phicrit}
    \varphi_c=\sqrt{6\,\alpha}\text{Tanh}^{-1}\left(\sqrt{\dfrac{V_{0}^{1/2}}{3\,\alpha\,\lambda_c}}\right).
\end{align}
\textcolor{black}{For simplicity, we fix $\alpha = 1$, which corresponds to the Starobinsky limit within the broader $\alpha$-attractor framework~\cite{ Kallosh:2013yoa, Starobinsky:1980te}. This choice yields inflationary predictions in excellent agreement with current CMB observations~\cite{Planck:2018jri, BICEP:2021xfz}}.
\begin{itemize}
    \item The amplitude of the scalar power spectrum, denoted as $A_s(k_0)$, with a specific value of $2.137\times 10^{-9}$ (as given in Eq. \eqref{curv})
    \item The number of e-folds, represented as $N_0$ in Eq. \eqref{Ngen}, is set to be 55 efolds.
     \item  The spectral index $n_{s}$ is set $n_{s}=0.96489$  central value.
\end{itemize}
These constraints play a crucial role in determining the predictions of the model. With these constraints imposed, the parameters $\kappa_c$, \(V_0\), $f$, and $m$ are varied. The results of our numerical calculations are presented in Fig \ref{results1}, where the behavior of various parameters is shown in the $\lambda_c-r$ plane. In our analysis, the scalar spectral index is fixed at the central value of Planck’s bounds, $n_s = 0.96489$ \cite{Planck:2018jri}. The boundary curves in Fig \ref{results1} represent the range $2.3 \times 10^{-7} \leq \lambda_c \leq 1$ and the decay constant $10^{-2} \leq f \leq 10^{2}$. 

The top panel in Fig. \ref{results1} shows the variation of the decay constant $f$ (left) and the variation of the axion mass $m$ (right) in the $\lambda_c-r$ plane. Similarly, the bottom panel shows the variations of $V_0$ in the $\lambda_c-r$ plane. Using the leading-order slow-roll approximation, we obtain the following analytical expressions for $r$. 
\begin{equation}
   r= \frac{1}{24\,\pi^2}
\left( \frac{16V(\varphi_0)}{A_{s}(k_0) }\right)\simeq \frac{1}{24\,\pi^2}
\left( \frac{16(V_0+m^2f^2)}{A_{s}(k_0) }\right),
\end{equation}
which explains the behavior of tensor to scalar ratio r in the  $\lambda_c-r$ plane. It can readily be checked that for $V_0\sim 5.88 \times10^{-11}$, $m\sim 3.39 \times 10^{-6}$ and $f\sim 0.884$, we obtain $r \approx 0.00204264 $. These approximate values are very close to the actual values obtained in the numerical calculations. The above equation therefore provides a valid approximation of our numerical results. Primordial gravity waves, associated with the tensor-to-scalar ratio \( r \), are expected to be measured with higher accuracy in the next-generation experiments such as  LiteBIRD \cite{LiteBIRD:2022cnt} and CMB-S4\cite{CMB-S4:2016ple} e.t.c. Large tensor modes are easily obtained, approaching observable values potentially measurable by these upcoming experiments. For the scalar spectral index $n_s$ fixed at Planck’s central value $(0.96489)$ \cite{Planck:2018jri, Planck:2018vyg}, we obtain the following ranges of parameters:
\begin{gather}
	\nonumber
	(10^{-2} \lesssim f\lesssim 10^{2}) , \\ \nonumber
	(1.5 \times 10^{-11} \lesssim m \lesssim 6.3 \times 10^{-6}) ~ , \\ 
 (2.3 \times 10^{-7} \lesssim \lambda_c =\lambda^2/\kappa\lesssim 1) ~ , \\ \nonumber
	(1.33 \times 10^{-20} \lesssim V_{0}=\kappa^2 M^4 \lesssim 1.8 \times 10^{-9}) ~ . \\ \nonumber
\end{gather}
We turn now to Scalar Perturbations and Primordial Black Hole Formation  and compute the relevant observables.

\section{Scalar Perturbations and Primordial Black Hole Formation}\label{sec3}
Primordial Black Holes (PBHs) form through the gravitational collapse of over-dense regions, triggered when their density contrasts upon horizon re-entry during radiation domination exceed a critical threshold \cite{Matarrese:1997ay}. These over-dense regions originate from primordial curvature perturbations generated during inflation. To produce a substantial abundance of PBH dark matter (DM), it is necessary to significantly amplify the amplitude of the power spectrum by approximately seven orders of magnitude \cite{PBH_dark}. These amplified perturbations enhance the scalar power spectrum at small scales, leading to the collapse of large density fluctuations upon horizon re-entry, thus forming PBHs. The presence of a linear term in the potential \cref{canonpoten} causes the field $\psi$ to deviate from $\psi=0$, but will be displaced depending upon the sign of the linear term coefficient $b$ \cite{Braglia:2022phb,Afzal:2024hwj,Afzal:2024xci}. This mechanism helps mitigate the formation of unnecessary topological defects and regulates the peak of the power spectrum at small scales to prevent excessive PBH production.  Before delving into the discussion on scalar perturbations, it is essential to elucidate the generation and evolution of scalar fields during inflation.
\begin{table}[h!]
	\centering
	{
\begin{tabular}{ |l|p{2.5cm}|p{2.5cm}|p{2.5cm}|p{3cm}| p{2.5cm}|}
\hline
\rowcolor{lightgray} \multicolumn{5}{|c|}{\textbf{Benchmark Points}} \\
\hline
\rowcolor{yellow!50}
Parameters& BP-1  & BP-2 & BP-3 & BP-4 \\
\hline
$V_{0}$ & $5.84\times10^{-11}$ & $5.87\times10^{-11}$  & $1.09 \times 10^{-10}$ & $9.991 \times 10^{-11}$\\
$m/m_\text{Pl}$ &$4.83\times 10^{-5}$ & $4.8\times 10^{-6}$ & $2.77 \times 10^{-6}$ & $2.37 \times 10^{-6}$\\
$f/m_\text{Pl} $  & 100 & 10  & 1 & 0.884 \\
$b$  &$-7.7\times 10^{-6}$  & $-7.7\times 10^{-6}$ & $-7.7\times 10^{-6}$ & $-8.5\times 10^{-6}$ \\
$\lambda_{c}$ & $10^{-4}$ &  $10^{-5}$ & $\times 10^{-5}$ & $4.9 \times 10^{-5}$\\
$\sqrt{\alpha}/m_\text{Pl}$ & 1 & 1 & 1 & 1\\
$\varphi_i/m_\text{Pl}$ & $5.37$ &   $5.35$  & 2.81 & 2.25  \\
$\psi_i/m_\text{Pl}$ & 0 &  0    & 0  & 0\\
$\varphi_c/m_\text{Pl}$ & $3.9 \times 10^{-1}$& $1.33 $& 0.77 & 0.651\\
$N_k$ & 62 & 58& 55 & 53 \\
$n_s$ & 0.96489 & 0.96489    & 0.96489  & 0.96489 \\
$r$ & 0.0038  & $0.0037$&  0.0034 &  0.0032\\
$\varphi_e/m_\text{Pl}$ & $2.3 \times 10^{-10}$  & $1 \times 10^{-8}$& $1.3 \times 10^{-4}$ & $1 \times 10^{-5}$ \\
$\psi_e/m_\text{Pl}$ & 2.18    & 2.18 & 2.18 & 2.18  \\
$f_\text{PBH}$ & $0.907$  & $0.4$& $0.0861$ & $0.0278879$ \\
$M_\text{PBH}$ & $5.66\times 10^{-16} $  &$ 4\times 10^{-12}$ & $5.79\times10^{-6}$ & $0.0015$  \\
$T_\text{reh}/m_\text{Pl}$ & $2.2\times 10^{-12} $  &$ 7.54\times 10^{-13}$ & $3.3\times10^{-12}$ & $2.96\times10^{-12}$  \\

\hline
\end{tabular}}
	\caption{\it  Benchmark points as shown in Figures.
 }
 \label{parmsets}
\end{table} 
\subsection{Scalar Spectra}

\begin{figure}[t]
    \centering
    \includegraphics[width=0.48\linewidth,height=6.7cm]{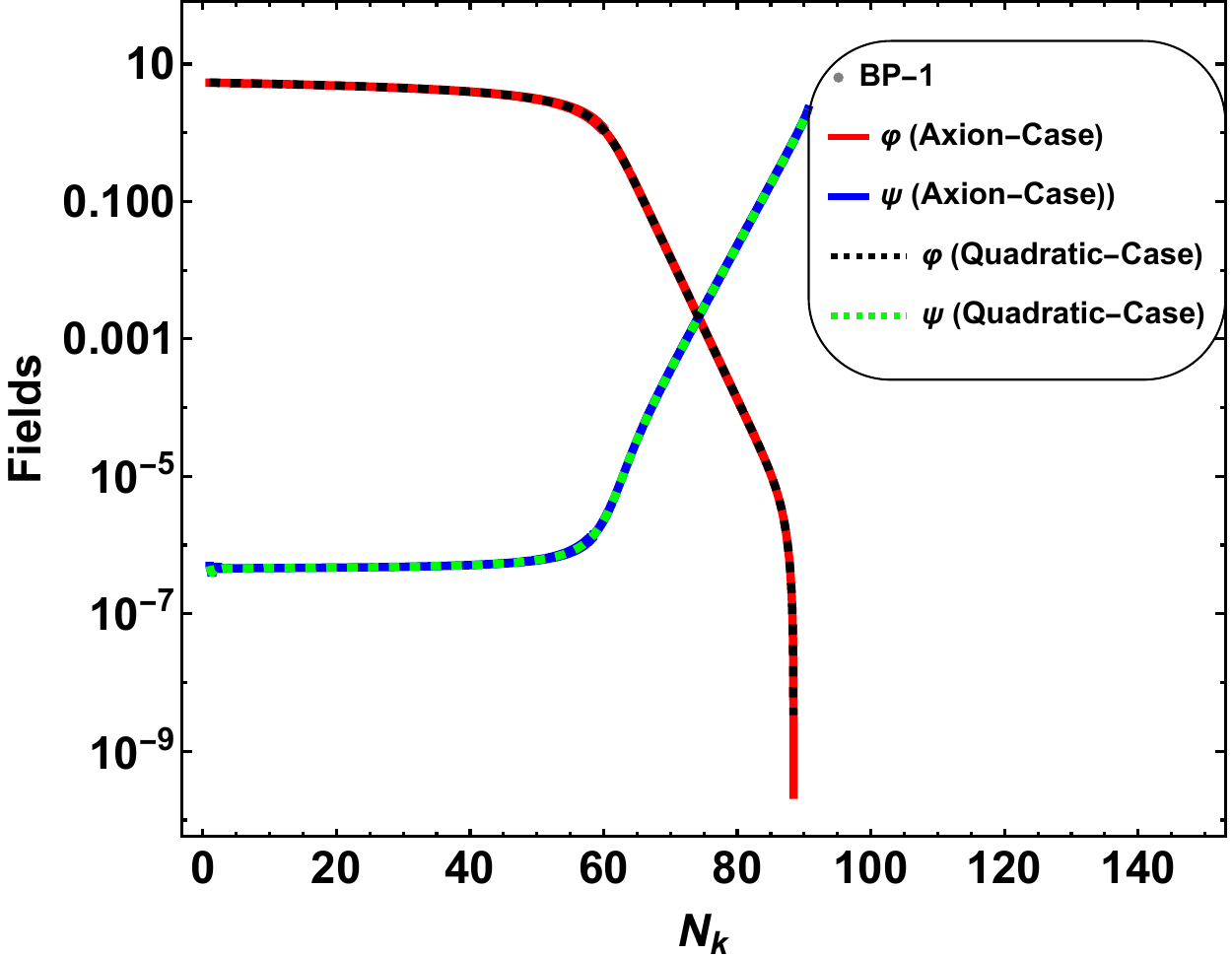}
    \quad
    \includegraphics[width=0.48\linewidth,height=6.7cm]{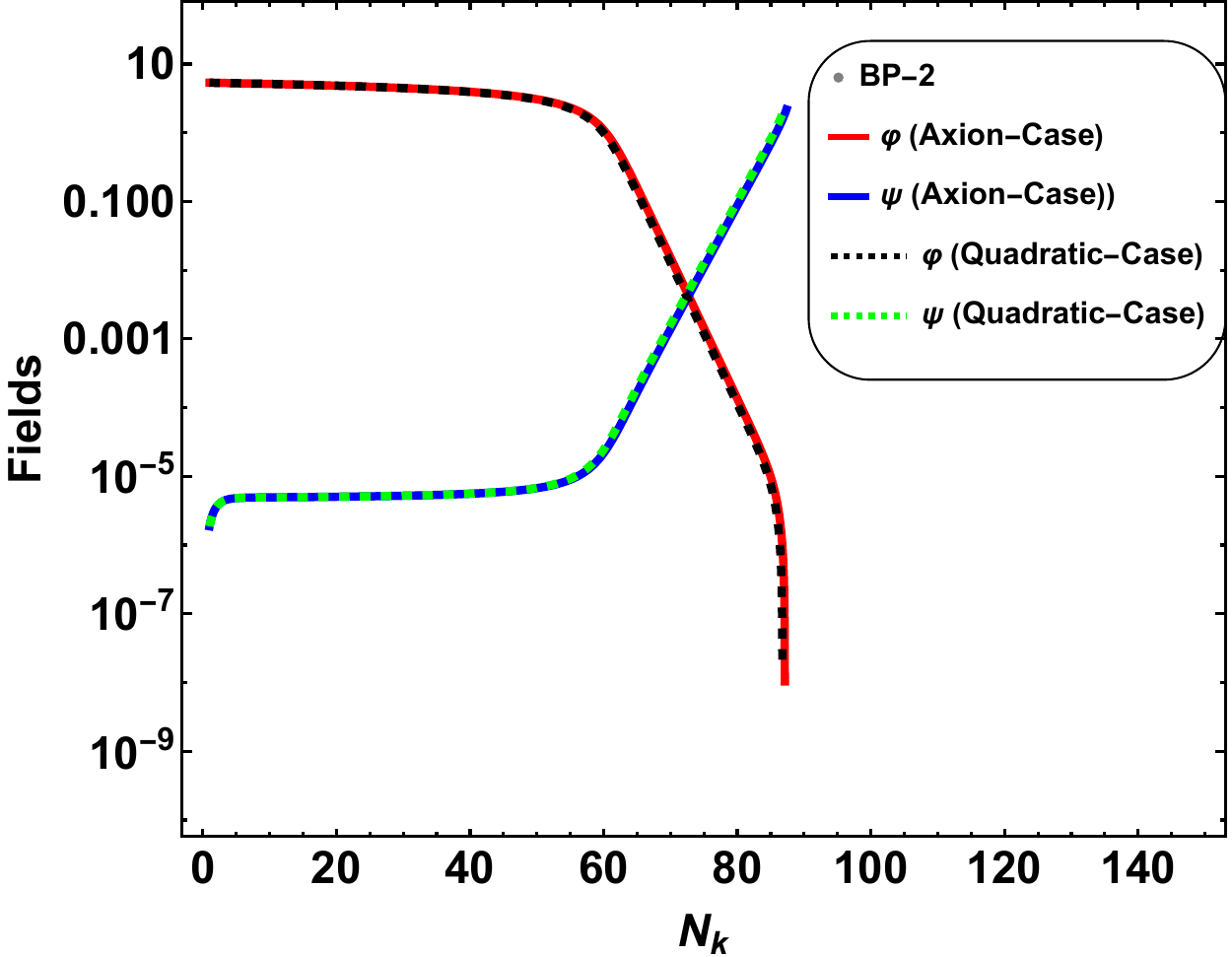}
      \quad
       \includegraphics[width=0.48\linewidth,height=6.7cm]{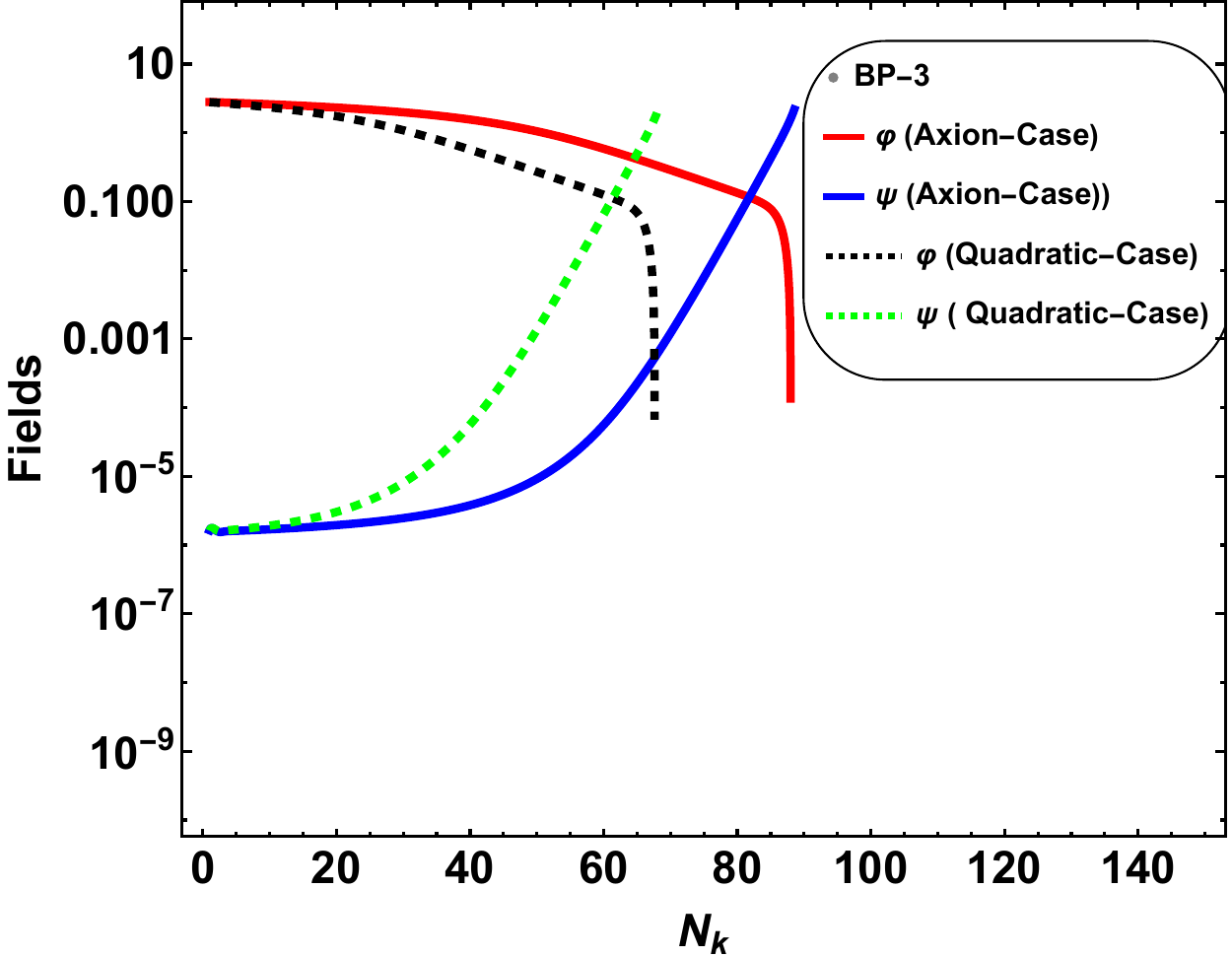}
      \quad
      \includegraphics[width=0.48\linewidth,height=6.7cm]{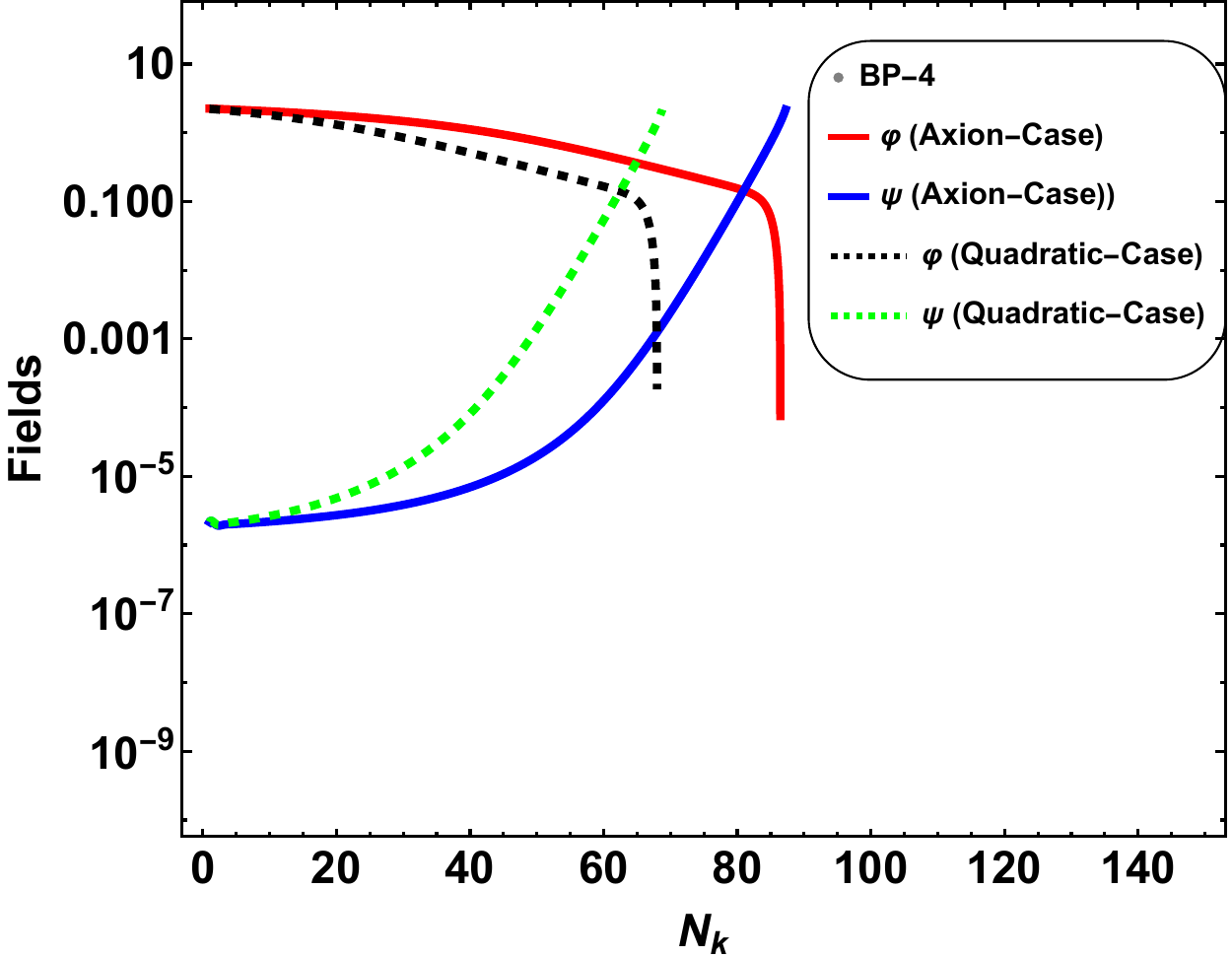}
      \quad
       \caption{\it  We compare the field evolution for axion and  quadratic potentials with the number of e-folds from the pivot scale to the end of inflation . For large \( f \) values, both potentials converge, with more evolution in the axion field $\varphi$  while the \( \psi \) field remains the same in both cases. As \( f \) decreases, the divergence in field evolution increases, showing the opposite trend. This is assessed by solving the background equation \cref{bac} using the potential in equation \cref{canonpoten} for the benchmark points in Table \ref{parmsets}.}
       \label{fig:phipsi}
\end{figure}

The background equations of motion in the number of e-fold times are given by \cite{Clesse:2013jra},
\begin{align}
\label{bac}
	\varphi^{''}+\left(\dfrac{H^{'}}{H}+3\right)\varphi^{'}+\dfrac{V_\varphi}{H^2}=0,\,\,\,\,\,\,\,\,\,\,\,\,\,\,
	\psi^{''}+\left(\dfrac{H^{'}}{H}+3\right)\psi^{'}+\dfrac{V_{\psi}}{H^2}=0,
\end{align}
where $H$, the Hubble rate is defined to be, $H^2=2\,V/(6-\varphi^{'2}-\psi^{'2})$.
Here, prime is the derivative with respect to the number of e-folds and $V_\xi=dV/d\xi$ where $\xi=\{\varphi,\psi\}$. The evolution of the fields is depicted in \cref{fig:phipsi}, showing the calculations up to the end of inflation, which occurs when $\epsilon_V = 1$. We used four different benchmark points as shown in \cref{parmsets}. The field evolution is illustrated in two cases: solid lines for the axionic potential, which is our main focus, and dotted lines for the  quadratic model with a tree-level potential as discussed in \cite{Afzal:2024xci}. In our analysis, we observe that for large values of $f$, both models converge, indicating similar behavior in their field evolution. However, the inflaton field in the axion case demonstrates slightly more evolution compared to the chaotic model, while the waterfall field evolution remains consistent across both models.

As we decrease the value of $f$, the differences between the models become more pronounced. The axion model shows an extended evolution of the waterfall field, leading to a larger number of e-folds compared to the  quadratic model. This divergence highlights the unique characteristics of the axionic potential, particularly in scenarios with smaller $f$ values. Consequently, at smaller $f$, the two models exhibit distinct behaviors, with the axion model providing a richer structure in its field evolution.

Moving forward, we focus on the axionic potential to discuss its implications in early universe cosmology, such as primordial black hole production and secondary gravitational waves. We delve into the scalar perturbations in detail, particularly in the context of multifield (hybrid) model. The Scalar perturbations in the Friedmann–Lemaître–Robertson–Walker (FLRW) metric, expressed in the longitudinal gauge, can be described as follows:
\cite{Clesse:2013jra};
\begin{align}
	ds^2=a(\tau)^2\left[(1+2\,\Phi_{\text{B}})d\tau^2+\left[(1-2\,\Psi_{\text{B}})\delta_{ij}+\dfrac{h_{ij}}{2}\right]dx^idx^j\right],
\end{align}
where $\Phi_{\text{B}}$ and $\Psi_{\text{B}}$ are the Bardeen potentials, $h_{ij}$ is the transverse-traceless tensor metric perturbation i.e. $h_i^i= 0 = h^j
_{i,j}$. The conformal time $\tau$, is related to cosmic time, $dt=a\,d\tau$, $a$ being the scale factor. Working in the conformal Newtonian gauge we set, $\Phi_{\text{B}}=\Psi_{\text{B}}$. Following the dynamics given in \cite{Ringeval:2007am, Clesse:2013jra} to calculate the power spectrum,
the Klein-Gordon equation to evaluate scalar perturbations is given by,
\begin{align}
	\delta \xi_i^{''}+(3-\epsilon)\delta \xi_i^{'}+\sum_{j=1}^{2}\dfrac{1}{H^2}V_{\xi_i \xi_j}\delta \xi_j+\dfrac{k^2}{a^2H^2}\delta \xi_i=4\Phi^{'}_{\text{B}}\,\xi^{'}_i-\dfrac{2\,\Phi_{\text{B}}}{H^2}V_{\xi_i}.
\end{align}
Here $\xi$ with subscript $(i, j)$ refers to the fields ($\varphi$, $\psi$), $k$ is the comoving wave vector, the equation of motion for $\Phi_{\text{B}}$ is given by,
\begin{align}
	\Phi^{''}_{\text{B}}+(7-\epsilon)\,\Phi^{'}_{\text{B}}+\left(\dfrac{2\,V}{H^2}+\dfrac{k^2}{a^2H^2}\right)\Phi_{\text{B}}+\dfrac{V_{\xi_i}}{H^2}\,\delta \xi_i=0.
\end{align} 


\begin{figure}[htbp]
    \centering
    \includegraphics[width=0.48\linewidth,height=6.5cm]{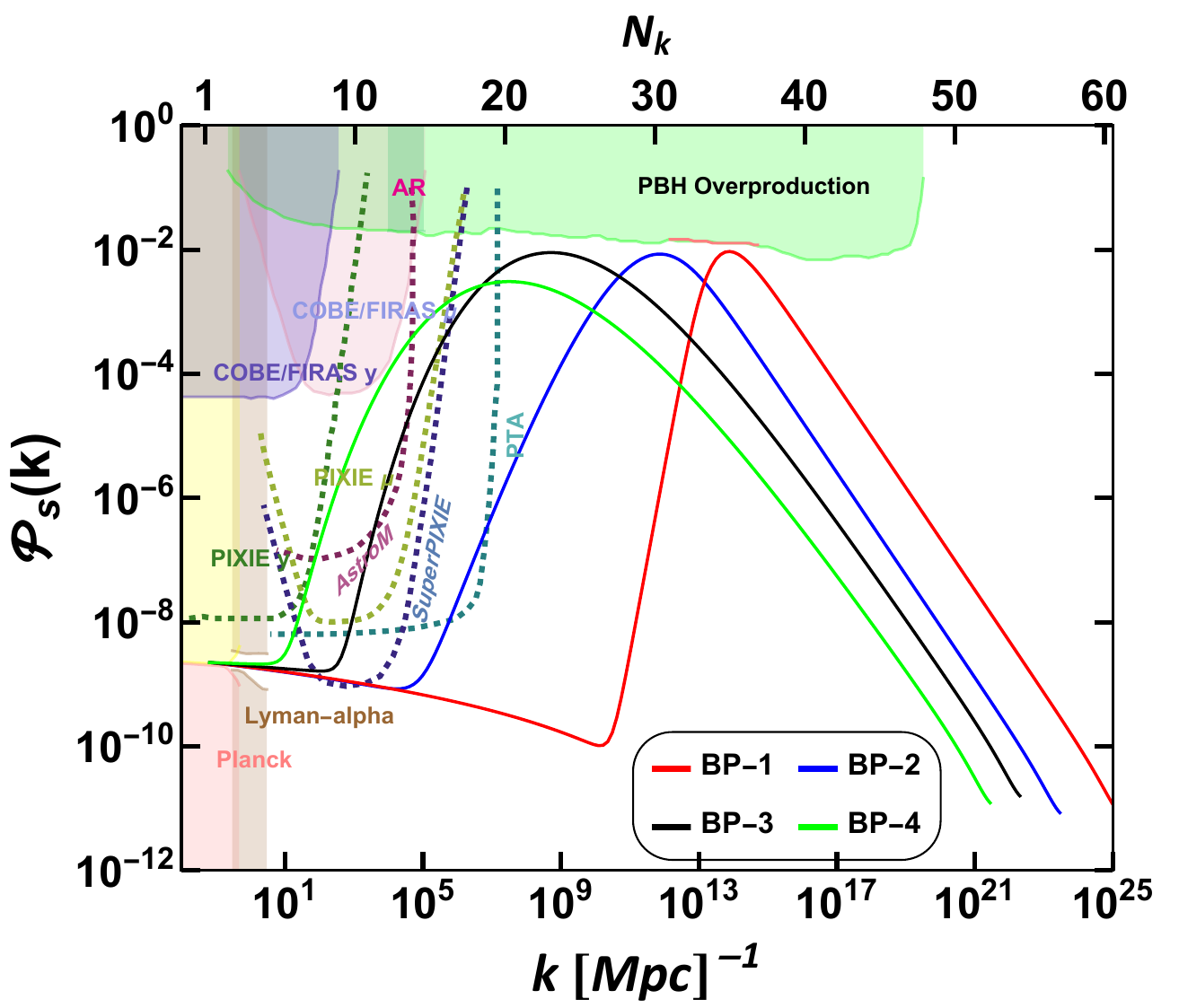}\\
    \includegraphics[width=0.48\linewidth,height=6.5cm]{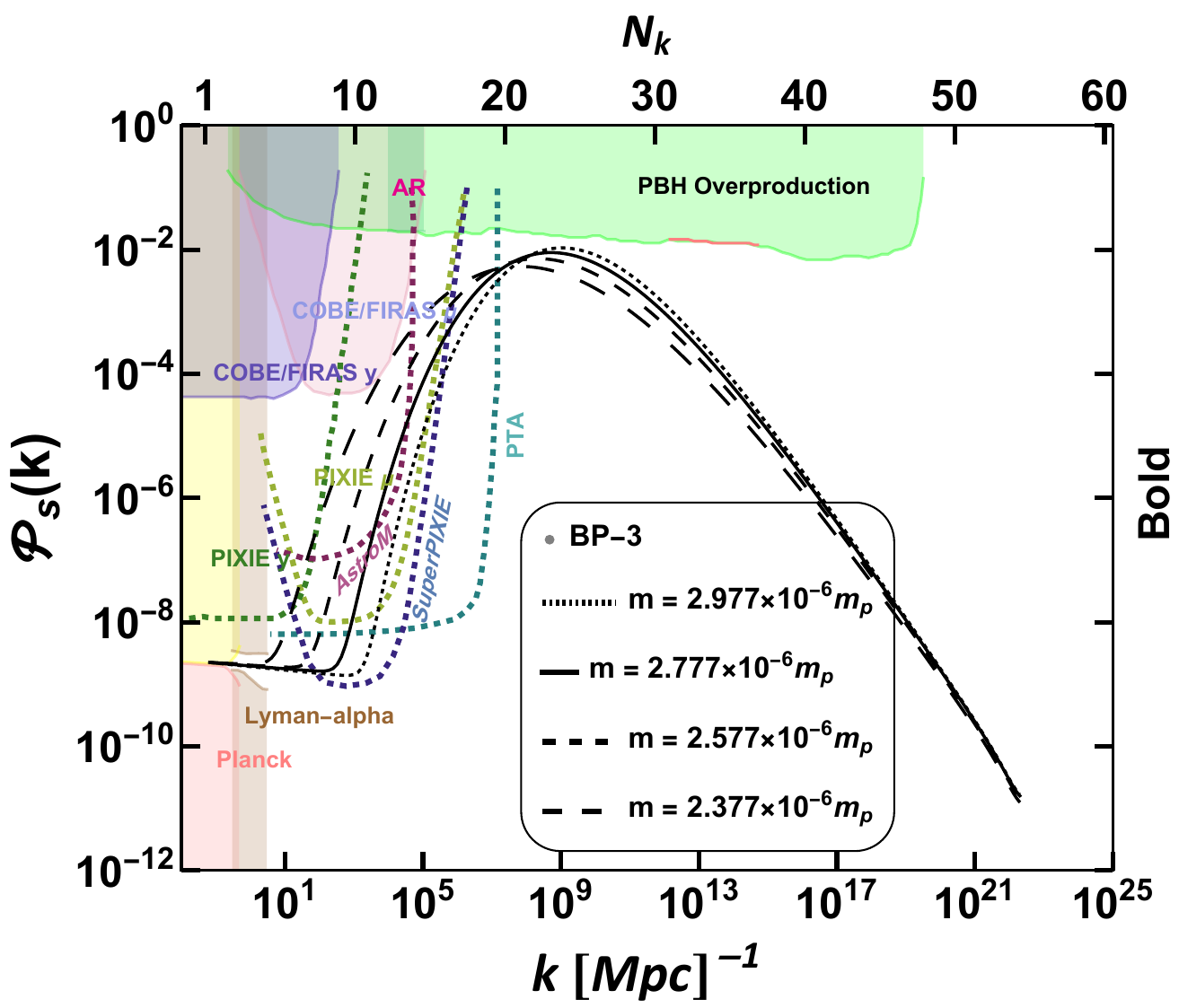}
      \quad
       \includegraphics[width=0.48\linewidth,height=6.5cm]{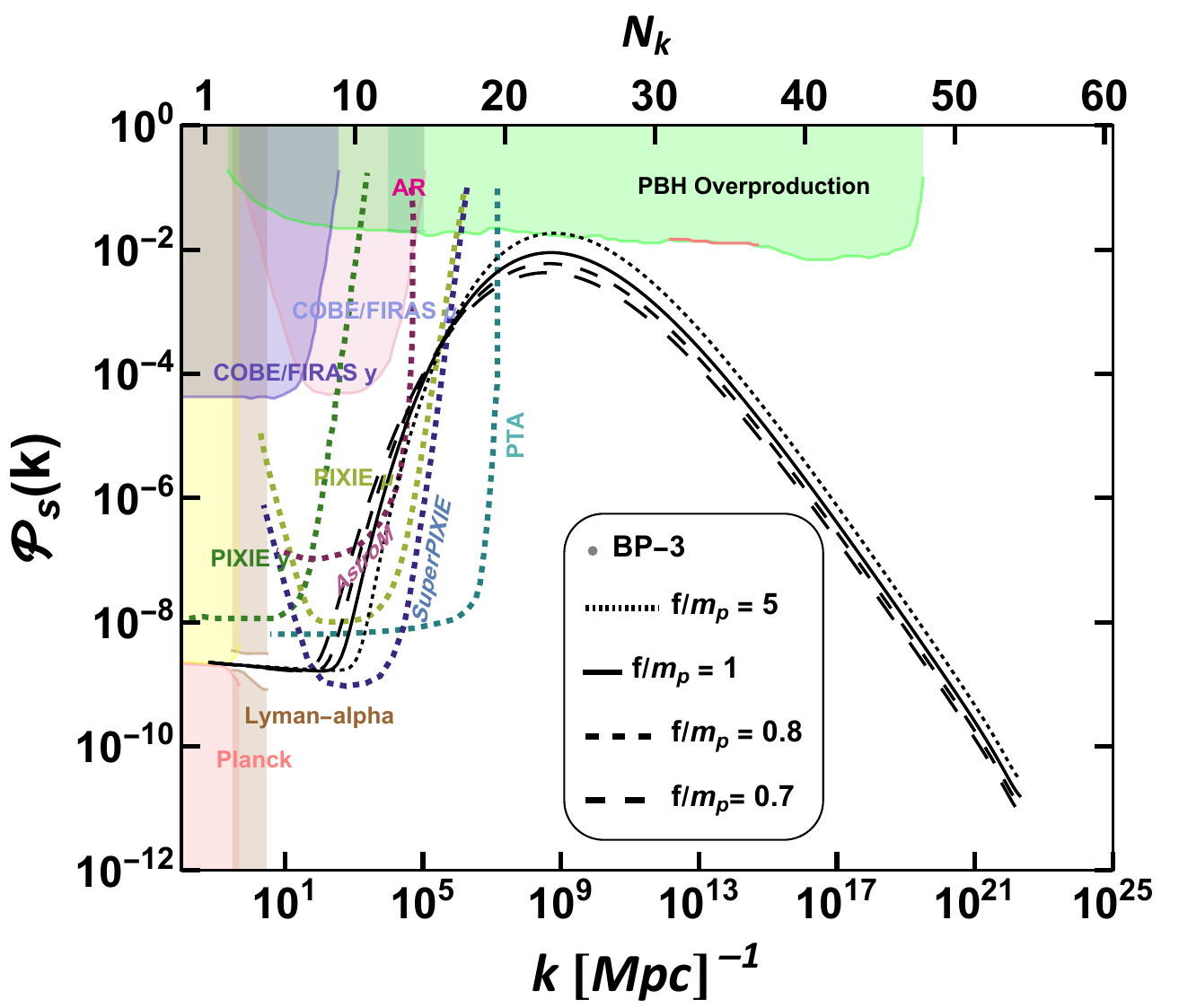}
      \quad
      \includegraphics[width=0.48\linewidth,height=6.5cm]{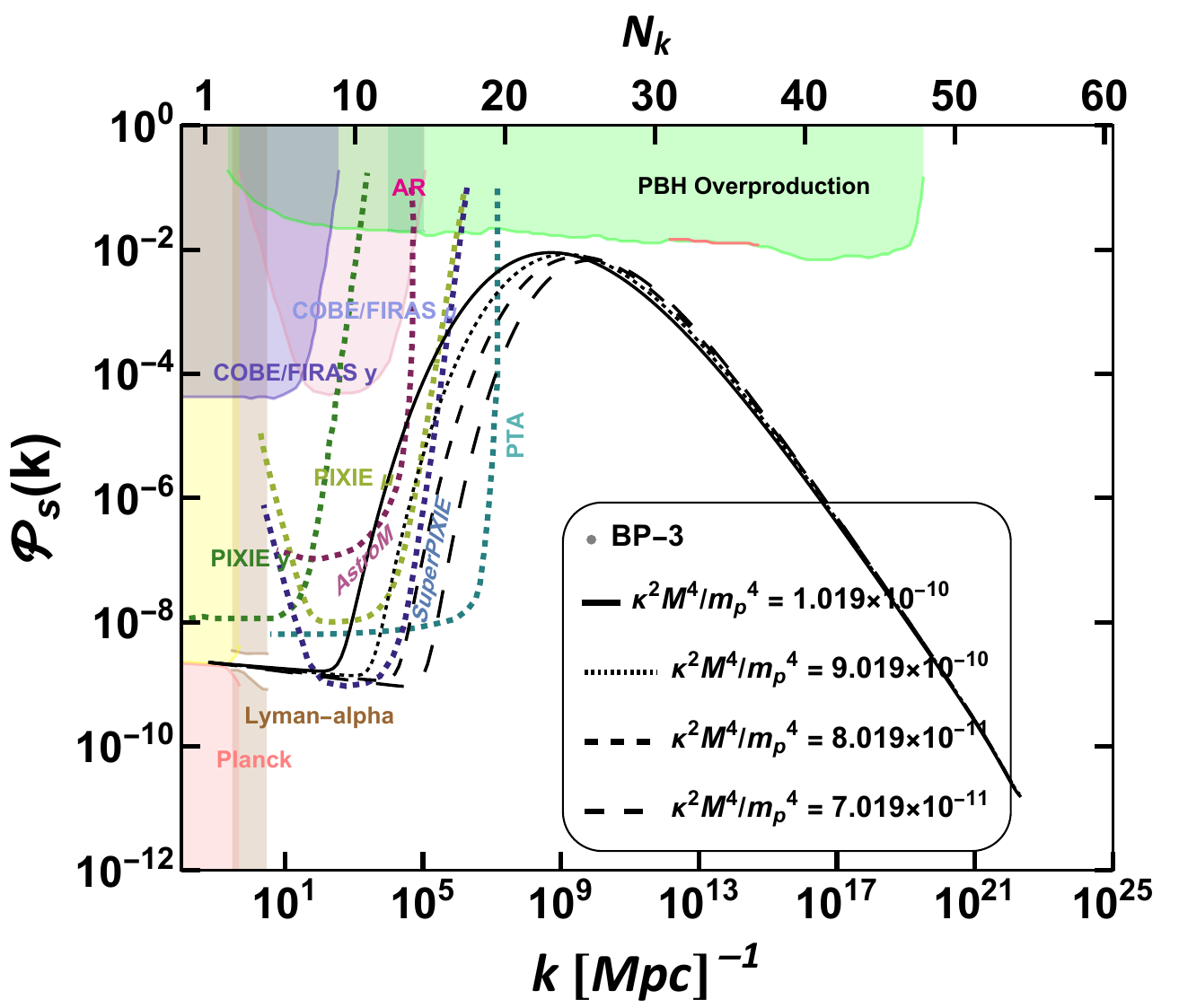}
      \quad
      \includegraphics[width=0.48\linewidth,height=6.5cm]{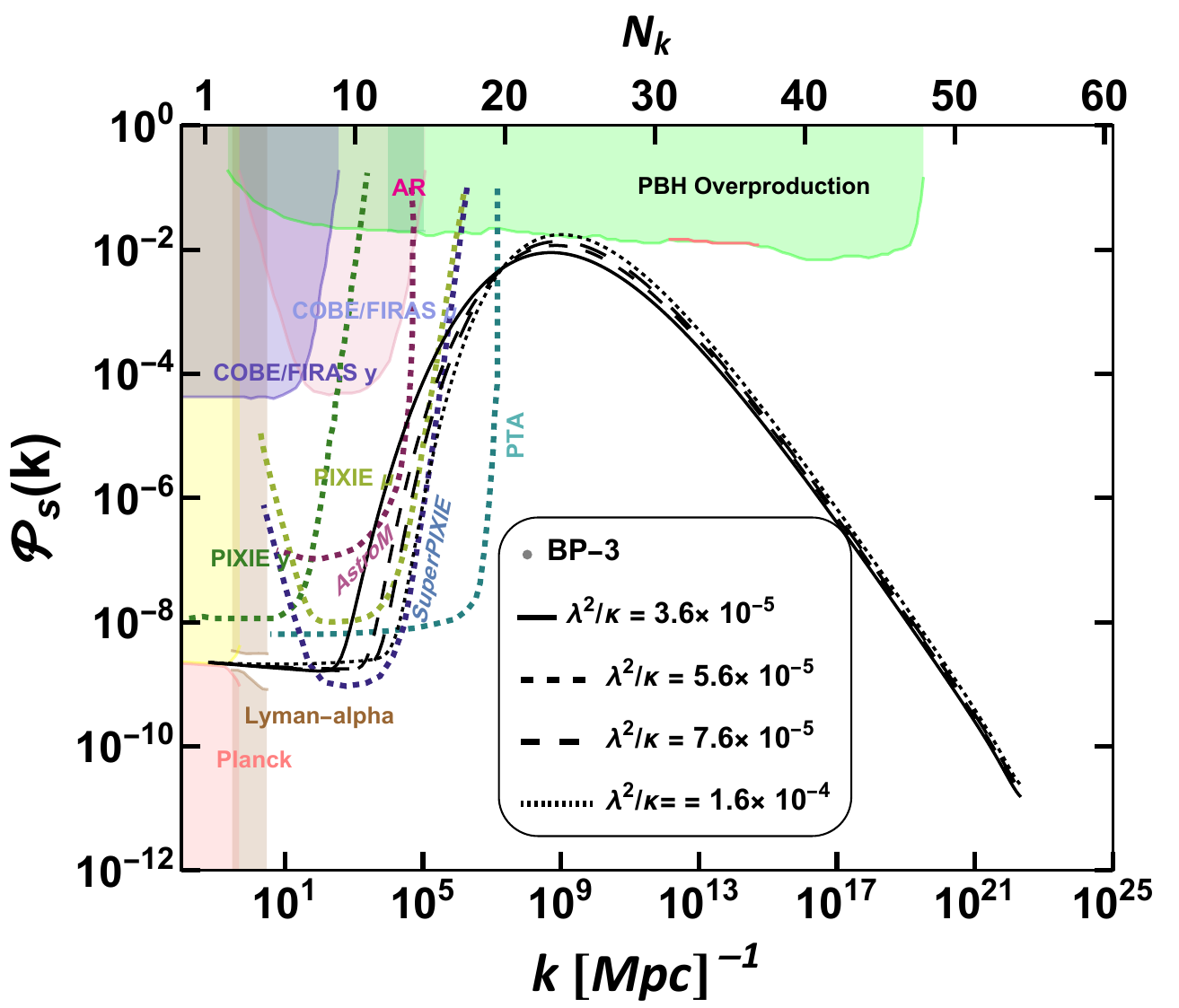}
      \quad
       \caption{\it  Power spectrum from the pivot scale to the end of inflation, obtained by solving the exact scalar perturbation equations for BP as given in Table \ref{parmsets}. The shaded area represents constraints from current (solid line) and and future (dashed line) experiments.}
        \label{fig:PS_k}
\end{figure}

To solve the equation of motion for $\Phi_{\text{B}}$, we consider the background dynamics in the context of the comoving wave vector $k$ and the fields $\xi = (\varphi, \psi)$. The term $\epsilon$ represents the slow-roll parameter, and $H$ is the Hubble parameter during inflation. The first term in the equation governs the damping due to expansion, while the second and third terms capture the potential and the spatial curvature's effects on the perturbation dynamics, respectively. The coupling between the perturbation $\Phi_{\text{B}}$ and the fields is encapsulated in the term $\dfrac{V_{\xi_i}}{H^2}\,\delta \xi_i$, which links the evolution of $\Phi_{\text{B}}$ with the fluctuations in the fields $\xi$. The initial conditions (i.c) for field perturbations in e-fold time are given as,
\begin{align}
	\delta \xi_{i,\text{i.c}}=\dfrac{1}{a_{\text{i.c}}\sqrt{2 k}},\,\,\,\,\,\,\,\,\,\,\,\,\,\,
	\delta \xi_{i,\text{i.c}}^{'}=-\dfrac{1}{a_{\text{i.c}}\sqrt{2 k}} \left(1+\iota\dfrac{k}{a_{\text{i.c}}H_{\text{i.c}}}\right).
\end{align}
The initial conditions for the Bardeen potential and its derivative are given by,
\begin{align}
	\Phi_{\text{B,i.c}}=\sum_{i=1}^{2}\dfrac{\left(H_{\text{i.c}}^2 \xi_{i,\text{i.c}}^{'}\delta \xi_{i,\text{i.c}}^{'}+\left(3H_{\text{i.c}}^2 \xi_{i,i.c}^{'}+V_{\xi_i,\text{i.c}}\right)\delta \xi_{i,\text{i.c}}\right)}{2H_{\text{i.c}}^2\left(\epsilon_{\text{i.c}}-\dfrac{k^2}{a_{\text{i.c}}^2H^2_{\text{i.c}}}\right)},\,\,\,\,
	\Phi^{'}_{\text{B,i.c}}=\sum_{i=1}^{2}\dfrac{\xi_{i,\text{i.c}}^{'}\delta \xi_{i,\text{i.c}}}{2}-\Phi_{\text{B,i.c}}.
\end{align}
The scalar power spectrum $\mathcal{P}_{\zeta}(k)$ is defined as,
\begin{align}
\label{Powspec}
	\mathcal{P}_{\zeta}(k)=\dfrac{k^3}{2\pi^2}\left|\zeta\right|^2=\dfrac{k^3}{2\pi^2}\left|\Phi_{\rm{B}}+\dfrac{\sum_{i=1}^{2}\xi_i\delta \xi_i}{\sum_{j=1}^{2}\xi^{'2}_j}\right|^2.
\end{align}

\textcolor{black}{Equation \ref{Powspec} gives the exact expression for the power spectrum, which we calculate numerically and explain in more detail later. However to connect the scalar field fluctuations with observable quantities such as the curvature perturbation $\zeta$, using the $\delta N$ formalism \cite{Starobinsky:1985ibc, Sasaki:1995aw, Lyth:2005fi}. This approach is particularly powerful for multi-field inflationary models, where superhorizon evolution of perturbations and isocurvature effects can be significant.}

\textcolor{black}{The $\delta N$ formalism relates the curvature perturbation on uniform energy density hypersurfaces to the perturbation in the number of e-folds, $N$, between an initial flat hypersurface (typically chosen at horizon crossing) and a final uniform-density hypersurface:
\begin{align}
\zeta(\mathbf{x}) = \delta N = N(\xi_i(\mathbf{x})) - N(\bar{\xi}_i),
\end{align}
where $\xi_i$ represent the scalar fields (in our case, $\xi_i = \{\varphi, \psi\}$), and $\bar{\xi}_i$ denote their homogeneous background values. Expanding this in a Taylor series gives:
\begin{align}
\zeta = \sum_i N_{,\xi_i} \delta \xi_i + \frac{1}{2} \sum_{i,j} N_{,\xi_i \xi_j} \delta \xi_i \delta \xi_j + \cdots,
\end{align}
where $N_{,\xi_i} \equiv \partial N/\partial \xi_i$ and $\delta \xi_i$ are the field fluctuations evaluated at horizon crossing. To leading order, the curvature perturbation is approximately:
\begin{align}
\zeta \approx N_{,\varphi} \delta \varphi + N_{,\psi} \delta \psi.
\end{align}
In the early stages of inflation, if the waterfall field $\psi$ is heavy (i.e., $m_\psi^2 \gg H^2$), its perturbations are suppressed and the dynamics are effectively single-field, with $\zeta \approx N_{,\varphi} \delta \varphi$. However, as the system approaches the critical instability point $\varphi_c$, the effective mass of $\psi$ decreases, and the field becomes tachyonic. In this regime, $\delta \psi$ becomes dynamically relevant and can significantly affect the total curvature perturbation. The multifield nature of the model therefore leads to enhanced superhorizon evolution and nontrivial contributions from isocurvature perturbations.The power spectrum of $\zeta$ can then be expressed as:
\begin{align}
\mathcal{P}_{\zeta}(k) = \left( \frac{H_*}{2\pi} \right)^2 \left( N_{,\varphi}^2 + N_{,\psi}^2 \right),
\end{align}
where all quantities are evaluated at the time of horizon crossing for a given mode $k = aH$. The $\delta N$ formalism thus provides an efficient semi-analytical way to estimate the curvature power spectrum, which we validate through numerical integration of the perturbation equations.} Using the potential \ref{canonpoten}, we conducted an exact numerical calculation of the power spectrum from the pivot scale to the end of inflation. Figure \ref{fig:phipsi} displays this spectrum for the four benchmark points listed in Table \ref{parmsets}. \footnote{
{\color{black} Usually, the evolution of quantum fluctuations
is affected due to the back-reaction of long modes. Although, for detailed analysis one needs
to perform the lattice simulations which is beyond the scope of the present work. As a naive
criterium, which should anyway give a good estimate of when back-reaction is negligible, we
may see that the background energy density; $H^2 {\phi ^{\prime}}^2/2 + m^2 \phi ^2/2 << M ^4$, for the BPs we chose
in Table  \ref{parmsets}. 
This enables us to disregard the backreaction on metric as discussed in detail in Refs. \cite{Braglia:2022phb, Garcia-Bellido:1996mdl}}} It is important to remark that our proposed models are in complete consistency with the observable constraints  $n_s$ and $r$ released by the Planck collaboration in 2018 \cite{Planck:2018jri}. The top panel aggregates the spectrum plots for all benchmark points, while the bottom four panels illustrate how the spectrum changes with variations in the parameters of the potentials. For comparison, we also include current and future bounds from various observational and experimental sources, such as Planck \cite{Planck:2018jri}, Lyman-alpha forest data \cite{Bird:2010mp}, PIXIE \cite{A_Kogut_2011}, COBE/FIRAS \cite{Fixsen:1996nj}, Super-PIXIE \cite{Chluba:2019kpb}, AstroM\cite{VanTilburg:2018ykj}  and Pulsar Timing Arrays (PTA) \cite{Lee:2020wfn, NANOGrav:2023gor, NANOGrav:2023hvm}.

Additionally, at scales $10^{-4} \lesssim k/\text{Mpc}^{-1} \lesssim 1$, the power spectrum is constrained by the angular resolution of current CMB measurements. Inhomogeneities at these scales result in isotropic deviations from the usual black body spectrum, known as CMB spectral distortions \cite{Chluba:2012we}. These distortions are categorized into $\mu$-distortions, associated with chemical potential occurring at early times, and Compton $y$-distortions, generated at redshifts $z \lesssim 5 \times 10^4$. A $\mu$-distortion is associated with a Bose-Einstein distribution with $\mu \neq 0$. The most stringent constraints on spectral distortions come from the COBE/FIRAS experiment, which restricts $\lvert \mu \rvert \lesssim 9.0 \times 10^{-5}$ and $\lvert y \rvert \lesssim 1.5 \times 10^{-5}$ at 95\% confidence level \cite{Fixsen:1996nj}. We also present acoustic reheating (AR) constraints on the spectrum \cite{Nakama_2014}. Note that solid lines represent current experiments, while dashed lines represent future experiments. Future detectors like PIXIE can investigate distortions with magnitudes $\mu \lesssim 2 \times 10^{-8}$ and $y \lesssim 4 \times 10^{-9}$ \cite{A_Kogut_2011}.


\subsection{ Primordial Black Hole formation} 
{\color{black}In  multi-field hybrid inflation model, entropy perturbations arise naturally due to the field-space dynamics of \(\phi\) and \(\psi\). Near the critical point, tachyonic instability in \(\psi\) amplifies entropy perturbations, which sources curvature perturbations, enhancing \(\mathcal{P}_{\zeta}(k)\). This mechanism can lead to PBH formation while remaining consistent with isocurvature constraints. \cite{Braglia:2022phb, Ezquiaga:2017fvi}}
In this section, we explore how enhanced curvature perturbations can trigger the formation of Primordial Black Holes (PBHs) through gravitational collapse as they reenter the horizon. We delve into calculating the mass of PBHs and their fractional energy density abundances. Our analysis assumes that PBH formation occurs during the radiation-dominated epoch.

The mass of a PBH is contingent upon the size of the horizon at the moment when the perturbation, associated with the PBH, crosses the horizon. This relationship between the perturbation scale and the PBH mass at the time of formation is expressed as follows:
\begin{equation}\label{eq4_1}
M_\text{PBH}=\gamma M_{H,0}\Omega_{rad,0}^{1/2}\left(\frac{g_{*,0}}{g_{*,f}}\right)^{1/6}\left(\frac{k_0}{k_f}\right)^2,
\end{equation}
where $M_{H,0}=\frac{4\pi}{H}$ represents the mass of the horizon, $\Omega_{rad}$ is used to denote the energy density parameter for radiation, while $g_*$ represents the effective degrees of freedom. Furthermore, the $f$ and $0$ in the subscript correspond to the time of formation and today respectively. $\gamma$ represents the ratio between the PBH mass and the horizon mass, which is estimated as $\gamma\simeq3^{-3/2}$ in the simple analytical result \cite{Carr:1975qj}.

\begin{figure}[t]
    \centering
    \includegraphics[width=0.8\linewidth]{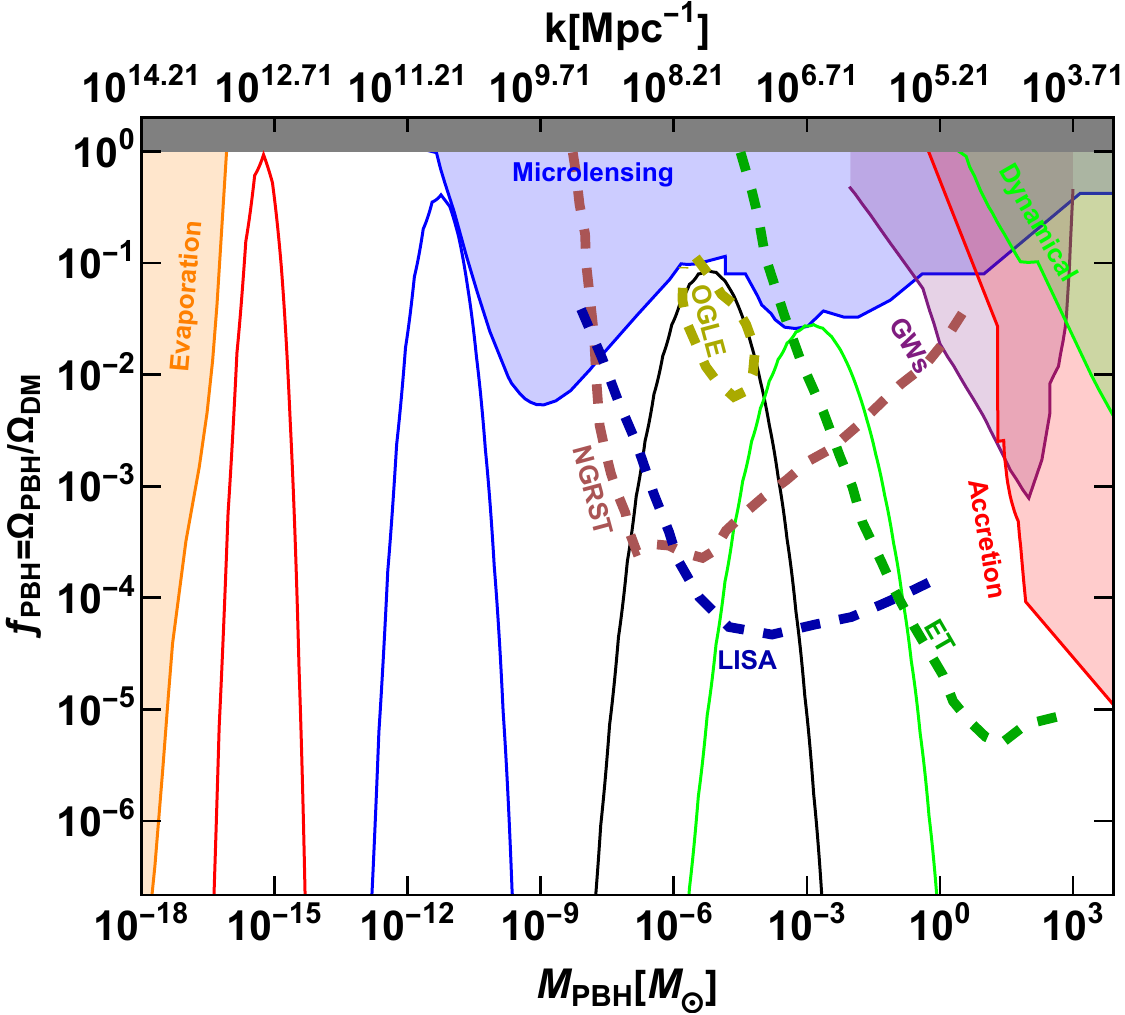}
    \caption{\it  Fraction of dark matter in the form of PBHs with
mass $M_{PBH}$. The shaded regions represent the observational constraints on the PBH abundance from various experiments (solid lines present and dashed for future), see the main text for the details.}
    \label{fig:PBHab}
\end{figure}
The energy density of the PBHs today can be obtained by redshifting that at the formation time, namely $\rho_\text{PBH,0}=\rho_\text{PBH,f}(a_\text{f}/a_0)^3\approx \gamma \beta \rho_\text{rad,f}(a_\text{f}/a_0)^3$, since the PBHs behave as matter. Here $\beta$ denotes the mass
fraction of Universe collapsing in PBH mass. The mass fraction $\beta$ is evaluated using the Press-Schechter method assuming that the overdensity $\delta$ follows a Gaussian probability distribution function. The collapse of PBH is determined by a threshold value denoted as $\delta_c$. Thus the mass fraction is given via the integral,
\begin{equation}
    \beta=\int_{\delta_c}^{\infty}d\delta\frac{1}{\sqrt{2\pi\sigma^2}}\exp\left(-\frac{\delta^2}{2\sigma^2}\right),
\end{equation}
where $\sigma$ is the variance of curvature perturbation that is related to the co-moving wavenumber \cite{Young}.
\begin{equation}
    \sigma^2=\frac{16}{81} \int^{\infty}_{0}  d\ln q (q k^{-1})^4 W^2 (q k^{-1}) \mathcal{P}_{\zeta}(q),
\end{equation}
where $W^2(q k^{-1})$ is a window function that we approximate with a Gaussian distribution function,
\begin{equation}
    W(q k^{-1})=\exp\left[-\frac{1}{2}(q k^{-1})^2\right].
\end{equation}
For $\delta_c$ we assume values in the range between $0.4$ and $0.6$ \cite{Musco:2020jjb,Escriva:2020tak,Escriva:2019phb,Musco:2018rwt} that corresponds to $\sigma^2\sim 10^{-3}-10^{-2}$. The total abundance $\Omega_\text{PBH}$ expressed in terms of $f_{\text{PBH}}$ which is given by,
\begin{align}
    f_{\text{PBH}}&\equiv\frac{\Omega_{\text{PBH,0}}}{\Omega_{\text{CDM,0}}}\nonumber\\
    &\approx\left(\frac{\beta(M)}{8.0\times10^{-15}}\right)\left(\frac{0.12}{\Omega_{\text{CDM,0}}h^2}\right)\left(\frac{\gamma}{0.2}\right)^{3/2}\nonumber\\&
    \times \left(\frac{106.75}{g_{*,f}}\right)^{1/4} \left(\frac{M_\text{PBH}}{10^{20} \text{g}}\right)^{-\frac{1}{2}},
\end{align}
where  $\Omega_{CDM,0}$ is the today's density parameter of the cold dark matter and $h$ is the rescaled Hubble rate today. The fractional abundance of the primordial black holes as a function of their mass is shown in Figure~\ref{fig:PBHab} We have used the benchmark point parameters of table~\ref{parmsets}. 

The shaded region in Figure~\ref{fig:PBHab}, shows the constraints on $f_{\rm PBH}$, with detailed constraints discussed in \cite{Green:2020jor,Saha:2021pqf,Laha:2019ssq,Ray:2021mxu}. The evaporation of PBHs via Hawking radiation imposes stringent constraints from sources such as CMB \cite{Clark:2016nst}, EDGES \cite{Mittal:2021egv}, INTEGRAL \cite{Laha:2020ivk,Berteaud:2022tws}, and Voyager \cite{Boudaud:2018hqb}. Constraints from 511 keV gamma rays \cite{DeRocco:2019fjq}, and the extragalactic gamma-ray background (EGRB) \cite{Carr:2009jm} are also considered. 

Microlensing observations, including those from HSC (Hyper-Supreme Cam) \cite{Niikura:2017zjd}, EROS \cite{EROS-2:2006ryy}, OGLE \cite{Niikura:2019kqi}, and Icarus \cite{Oguri:2017ock}, provide additional constraints. Various constraints arise due to modifications of the CMB spectrum from PBH accretion, as explored in \cite{Serpico:2020ehh} and \cite{Piga:2022ysp}. 

The mass range around $M_{\odot}$ is constrained by LIGO-VIRGO-KAGRA (LVK) observations of PBH-PBH mergers \cite{Franciolini:2022tfm,Kavanagh:2018ggo,Hall:2020daa,Wong:2020yig,Hutsi:2020sol,DeLuca:2021wjr,Franciolini:2021tla}. Future gravitational wave detectors like LISA and Einstein Telescope (ET) are expected to further limit PBH abundance, as shown in \cite{DeLuca:2021hde,Pujolas:2021yaw,Franciolini:2022htd,Martinelli:2022elq,Franciolini:2023opt,Branchesi:2023mws}, represented by dashed lines in the plot. Additionally, we show the projected sensitivity of the Nancy Grace Roman Space Telescope (NGRST) for microlensing \cite{DeRocco:2023gde}.

For the parameter sets provided in Table \ref{parmsets}, PBHs can account for the entire abundance of DM in BP-1. However, for BP-2, only a fraction of DM can be explained due to various observational constraints from different experiments, as indicated by the shaded regions in Figure \ref{fig:PBHab}. According to equation \ref{eq4_1}, $M_{\text{PBH}} \propto \kappa^{-2}$. Thus, to explain PBHs in PTAs, the power spectrum must be enhanced at low $\kappa$ and large $M_{\text{PBH}}$, which are constrained by microlensing effects. BP-3 and BP-4 can explain these regions, but the entire abundance of DM from PBHs cannot be observed in PTAs due to different constraints.   We also show future sensitivity reaches of the Nancy Grace Roman Space Telescope (NGRST) from micro-lensing, see ref. \cite{DeRocco:2023gde}.
Our predictions in Table \ref{parmsets} indicate that for BP-1, PBHs account for the entire abundance of dark matter. In contrast, BP-2 shows that up to $40 \%$ of dark matter can be explained by PBHs. For BP-3 and BP-4, PBHs account for a smaller fraction of dark matter, which can be tested in future experiments such as the Nancy Grace Roman Space Telescope (NGRST) and the Einstein Telescope (ET).


\section{Scalar-induced Gravitational Wave} \label{sec4}

The energy density of gravitational waves (GWs) within the subhorizon regions can be expressed as \cite{Maggiore:1999vm},
\begin{equation}\label{gw1}
    \rho_{\text{gw}}=\frac{\langle \overline{\partial_lh_{ij}\partial^lh^{ij}}\rangle}{16a^2},
\end{equation}
where the average over oscillations is denoted by the overline and $h_{ij}$ is the tensor mode. We can decompose $h_{ij}$ as,  
\begin{equation}\label{gw2}
    h_{ij}(t,\textbf{x})=\int \frac{d^3\textbf{k}}{(2\pi)^{3/2}}\left( h_\textbf{k}^{+}(t)e_{ij}^{+}(\textbf{k})+h_{\textbf{k}}^{\times}(t)e_{ij}^{\times}(\textbf{k})\right)e^{i\textbf{k}\cdot\textbf{x}},
\end{equation}
with $e_{ij}^{+}(\textbf{k})$ and $e_{ij}^{\times}(\textbf{k})$ being the polarization tensors and can be expressed as,  
\begin{align}
 e_{ij}^{+}(\textbf{k})&=\frac{1}{\sqrt{2}}\left(e_{i}(\textbf{k})e_{j}(\textbf{k})-\bar{e}_i(\textbf{k})\bar{e}_j(\textbf{k})\right),\\
  e_{ij}^{\times}(\textbf{k})&=\frac{1}{\sqrt{2}}\left(e_{i}(\textbf{k})\bar{e}_{j}(\textbf{k})-\bar{e}_i(\textbf{k})e_j(\textbf{k})\right),
\end{align}
where $e_{i}(\textbf{k})$ and $\bar{e}_{j}(\textbf{k})$ are orthogonal to each other. Using expression (\ref{gw2}) in (\ref{gw1}) gives,
\begin{equation}
    \rho_{\text{gw}}(t)=\int d\ln k\frac{\overline{P_h{(t,k)}}}{2}\left(\frac{k}{2a}\right)^2.
\end{equation}
Here, $P_{h}\equiv P_h^{+,\times}$ and is defined as,
\begin{align}
    \langle h_{\textbf{k}}^{+}(t)h_{\textbf{q}}^{+}(t)\rangle&=\frac{2\pi^2}{k^3}P_{h}^{+}(t,k)\delta^3(\textbf{k}+\textbf{q}),\\
     \langle h_{\textbf{k}}^{\times}(t)h_{\textbf{q}}^{\times}(t)\rangle&=\frac{2\pi^2}{k^3}P_{h}^{\times}(t,k)\delta^3(\textbf{k}+\textbf{q}).
\end{align}
We note that $P_h^{+}(t,k)=P_h^{\times}(t,k)$. The energy density parameter for gravitational waves is defined as follows,
\begin{equation}
    \Omega_{\text{gw}}(t,k)\equiv\frac{\rho_{\text{gw}}(t,k)}{\rho_{\text{crit}}}=\frac{1}{24}\left(\frac{k}{aH}\right)^2\overline{P_h(t,k)},
\end{equation}
\begin{figure}[htbp]
    \centering
    \includegraphics[width=0.8\linewidth]{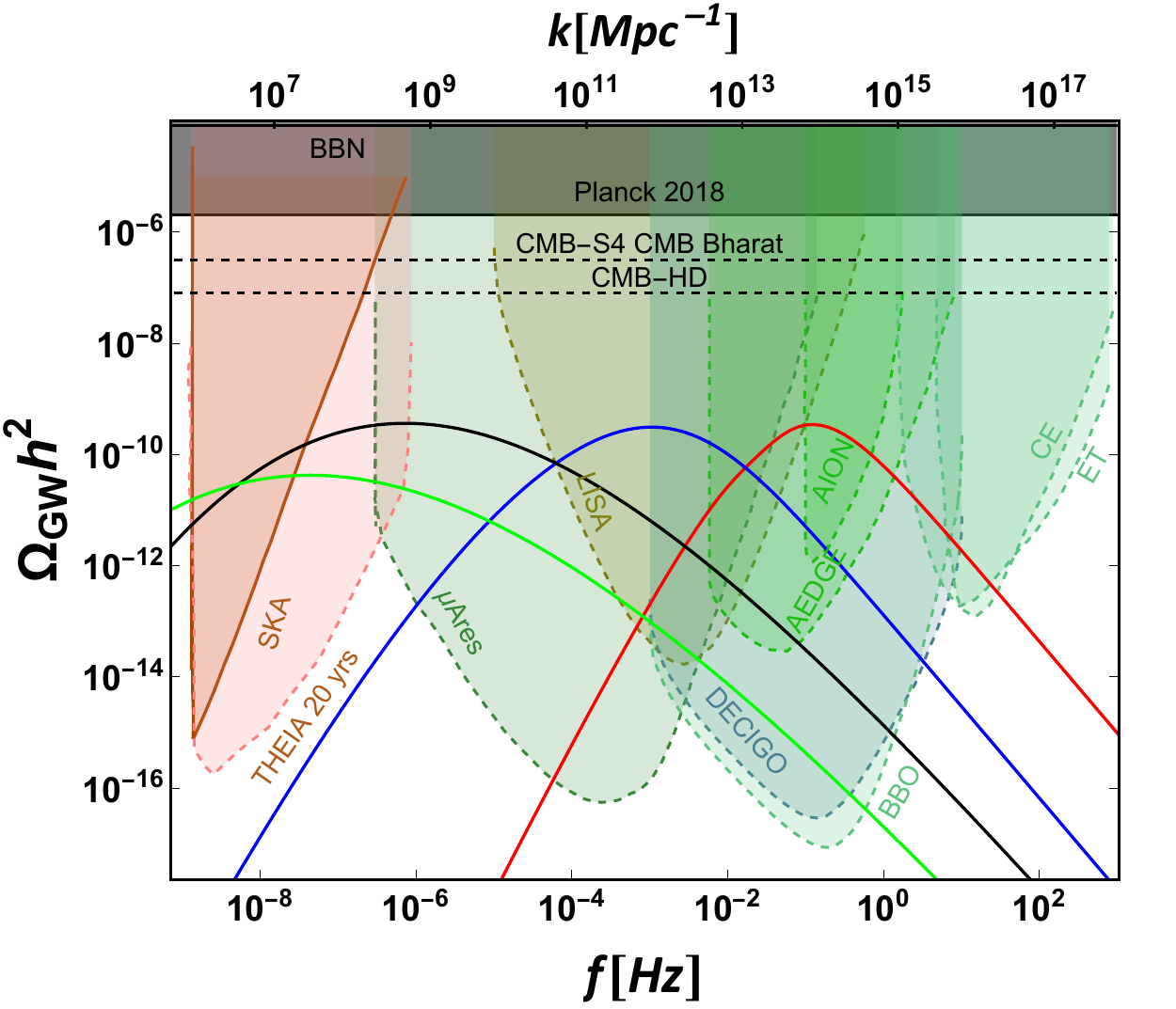}
    \caption{\it   The gravitational waves spectrum together with the sensitivity curves of the various existing and forthcoming GW experiments.}
    \label{fig:gws}
\end{figure}
where we have omitted the polarization index. To obtain the tensor power spectrum, we solve the following tensor perturbation equation \cite{Baumann:2007zm},
\begin{equation}
h^{\prime\prime}_{\textbf{k}}+2aHh_{\textbf{k}}^{\prime}+k^2h_{\textbf{k}}=4S_{\textbf{k}},
\end{equation}
where derivatives with respect to conformal time are denoted by primes. The source term $S_k$ is expressed as,
\begin{equation}
    \begin{split}
        S_k=\int&\frac{d^3q}{(2\pi)^{3/2}}q_iq_je_{ij}(\textbf{k})\biggr[2\Phi_{\textbf{q}}\Phi_{\textbf{k}-\textbf{q}} \\ & 
        +\frac{4}{3+3\omega}\left(\frac{\Phi^{\prime}_{\textbf{q}}}{H}+\Phi_{\textbf{q}}\right)\left(\frac{\Phi^{\prime}_{\textbf{k}-\textbf{q}}}{H}+\Phi_{\textbf{k}-\textbf{q}}\right) \biggr],
    \end{split}
\end{equation}
where $\omega$ is the equation of state. The generation of induced gravitational waves is assumed to occur during the radiation-dominated era. Thus, at the time of their generation, we have \cite{Kohri:2018awv},
\begin{equation}
   \begin{split}
        \Omega_{\text{gw}}(t_f,k) &= \frac{1}{12} \int_0^{\infty} dv \int_{|1-v|}^{1+v} du \left( \frac{4v^2-(1+v^2-u^2)^2}{4uv}\right)^2 \\ & \times \mathcal{P}_{\zeta}(ku) \mathcal{P}_{\zeta}(kv) \left(\frac{3(u^2+v^2-3)}{4u^3v^3} \right)^2 \\ &
        \times \left.\biggr[\left(\pi^2(-3+v^2+u^2)^2 \theta(-\sqrt{3}+u+v) \right) \right. \\ & \left.+\left(-4uv+(v^2+u^2-3)\text{Log} \left| \frac{3-(u+v)^2}{3-(u-v)^2} \right| \right)^2 \right].
    \end{split}
\end{equation}
We can determine the energy density parameter for the current time as follows \cite{Kohri:2018awv, Ando:2018qdb},
\begin{equation}
    \Omega_{\text{gw}}=\Omega_{rad,0}\Omega_{\text{gw}}(t_f),
\end{equation}
where we have multiplied $\Omega_{\text{gw}}(t_f)$ with today's radiation energy density parameter, $\Omega_{rad,0}$. Utilizing the numerically computed power spectrum of scalar perturbations, as depicted in Figure~\ref{fig:PS_k}, we derive the corresponding second-order gravitational wave (GW) spectrum. Our findings, are shown in Figure~\ref{fig:gws}, represent the energy density fraction of gravitational waves relative to the critical energy density, denoted as $\Omega_{\rm GW}$. Our analysis reveals that the predictions for all four benchmark points (BPs) intersect the sensitivity thresholds of upcoming experiments. Notably, the GW spectrum corresponding to BP-3 and BP-4 can be probed by space-based laser interferometers such as LISA \cite{LISA:2017pwj}, BBO \cite{Crowder:2005nr}, DECIGO \cite{Seto:2001qf}, Ground-based interferometers like the Einstein Telescope (ET) \cite{Punturo:2010zz}, Cosmic Explorer (CE) \cite{LIGOScientific:2016wof}, and the atomic interferometer AEDGE \cite{AEDGE:2019nxb}. Additionally, the low-frequency GW spectrum (1-10 nHz) for BP-3 and BP-4 can be investigated by planned pulsar timing arrays (PTAs) such as SKA \cite{Janssen:2014dka} and future plan experiments THEIA \cite{Garcia-Bellido:2021zgu}.

\section{Reheating Estimates} \label{sec5}
After the end of inflation, the inflaton system (axion), composed of two scalar fields, $\varphi$ and $\psi$, with mass $m_{\varphi}=m$ descends towards the minimum, undergoes damped oscillations around it, and eventually decays, initiating the process known as 'reheating'. The inflaton  may have couplings to the SM particles the interaction terms are read as

\begin{equation}
 \mathscr{L} \supset \frac{g_{\varphi \gamma \gamma}}{4} \varphi F_{\mu \nu} \tilde{F}^{\mu \nu}  +g_{\varphi \ell \ell} (\partial_\mu \varphi) (\bar{\ell} \gamma^\mu \gamma^5 \ell) +  \frac{g_{\varphi g g}}{4} \varphi G_{\mu \nu} \tilde{G}^{\mu \nu}.
\end{equation}
Where the first term represents the axion coupling to photons, the second term indicates its coupling to leptons, and the third term shows its interaction with gauge fields. However, the interaction between the axion and the Higgs boson can be seen through higher-dimensional operators in an effective field theory framework, which one can read as:
\begin{equation}
\mathscr{L} \supset \frac{g_{HH}}{f^2} (H^\dagger H) \partial_\mu \varphi \partial^\mu \varphi,
\end{equation}
The higgs interaction to axions is suppressed compared to other channels, which makes it relevant at higher energy scales. The corresponding total decay width is
\begin{equation}
\Gamma_{T} = \Gamma_{\varphi \to \gamma \gamma} + \Gamma_{\varphi \to \ell \ell} + \Gamma_{\varphi \to g g}    
\end{equation}
which can also read as,
\begin{equation}
\Gamma_{T} =  \frac{g_{\varphi \gamma \gamma}^2 m_\varphi^3}{64 \pi}   +\frac{g_{\varphi \ell \ell}^2 m_\varphi}{8 \pi} \left(1 - \frac{4m_\ell^2}{m_\varphi^2}\right)^{3/2}+ \frac{g_{\varphi g g}^2 m_\varphi^3}{64 \pi},
\end{equation}
where these couplings are defined as:
\begin{equation}
  g_{\varphi\gamma\gamma} = \frac{\alpha  C_\gamma}{2\pi f}, \quad g_{\varphi gg} = \frac{\alpha_s C_g}{2\pi f}, \quad g_{\varphi \ell\ell} = \frac{C_l m_\ell}{f}.  
\end{equation}
In the above expressions:
\begin{itemize}
    \item $\alpha$ is the fine-structure constant.
    \item $C_\gamma$, $C_G$ and $C_\ell$   is a model-dependent coefficient for simplcity we choose it equal 1.
    \item f is the axion decay constant.
    \item $\alpha_s$ is the strong coupling constant.
    \item $m_\ell$ is the mass of the lepton.
\end{itemize}
The reheating temperature can be estimated using the decay width $\Gamma_T$ of the inflaton field $\varphi$ into other particles. The relation is given by:
\begin{equation}
T_{\text{reh}} \approx \left( \frac{90}{\pi^2 g_*} \right)^{1/4} \sqrt{\Gamma_T}
\end{equation}
where $g_*=106.75$  for standard model, is the effective number of relativistic degrees of freedom. Table \ref{parmsets} shows the estimated reheating temperatures, $T_{\text{reh}}$ , for four different benchmark models. The values of $T_{\text{reh}}$ change based on the specific parameters used in each model. These parameters are constrained by observational data, such as the measurements from Planck, which include the scalar spectral index $n_s$ and the tensor-to-scalar ratio $r$. The typical reheating temperatures for the four benchmark points fall within the range of $( 10^6 - 10^7)$ GeV. This range is consistent with BBN constraints, supporting a viable cosmological framework for the formation of Primordial Black Holes PBHs and the generation of DM.

\section{Summary and Conclusion}\label{sec6}
In this study, we have explored the implications of axion inflation within hybrid inflationary backgrounds, focusing on the generation of primordial black holes and  secondary gravitational waves. We summarize the
 main findings of our analysis below
\begin{itemize}
    \item We adopted an $\alpha$-attractor scenario with a specific kinetic term configuration, as outlined in \cite{Kallosh:2022ggf}, to accurately predict cosmological parameters such as the scalar tilt $n_s = 0.9654$, consistent with the Planck 2018 bounds, along with a large tensor-to-scalar ratio $r \approx (10^{-11}-10^{-2})$, potentially measurable by forthcoming CMB experiments like LiteBIRD
    and CMB-S4 (see figure \ref{results1}). 

    \item  We compare the field evolution for axion and quadratic potentials with the number of e-folds from the pivot scale to the end of inflation for the bench mark points mention in Table \ref{parmsets}. For large $f$ values, both potentials converge, resulting in more evolution in the axion field $\varphi$ while the $\psi$ field remains unchanged in both cases. As $f$ decreases, the divergence in field evolution increases, demonstrating the opposite trend (see figure \ref{fig:phipsi}).

    \item We estimate the power spectrum across all $k$-values, providing constraints on model parameters such as axion mass and decay constant $f$ from measurements of CMB spectral distortions. Along the valley, when the inflaton field becomes smaller than the critical value $\varphi_c$, the effective mass squared of the waterfall field becomes negative. This induces tachyonic instability, causing the power spectrum to spike at small scales, as illustrated in the figure \ref{fig:PS_k}.

    \item   Our predictions indicate that for BP-1, PBHs account for the entire dark matter abundance, while BP-2 shows that up to $40\%$ of dark matter can be explained by PBHs due to observational limits (Figure \ref{fig:PBHab}). BP-3 and BP-4 match some regions but can't explain all dark matter in PTAs, which can be tested in future experiments such as the Nancy Grace Roman Space Telescope (NGRST) and the Einstein Telescope (ET).

    \item  Second-order tensor perturbations propagating as GWs with an amplitude $\Omega_{\text{GW}} h^2 $ verus $f$ as shown in \ref{fig:gws}. BP-1 can  be tested by can tested in LISA and  other experiments such as BBO, DECIGO, and ground-based interferometers like ET and Cosmic Explorer (CE), and it accounts for all dark matter. In contrast, BP-2 cannot be probed by ET and CE but can be tested by the other mentioned experiments at those frequencies.
    
 \item  The low-frequency GW spectrum (1-10 nHz) for BP-3 and BP-4 can be investigated by  planned pulsar timing arrays (PTAs) like SKA and future plan experiments THEIA . (see figure \ref{fig:gws}).

\item Finally, we discuss the reheating scenarios and possible decay channels of axions into Standard Model particles. The dominant contribution comes from the decay of axions into gluons. The reheating temperatures corresponding to all four benchmark points are summarized in Table \ref{parmsets}. The typical reheating temperatures are all within the range of $( 10^6 - 10^7)$ GeV consistent with BBN constraints.

\end{itemize}
As a future outlook, one could look to account for the potential influence of non-Gaussianity on PBH formation. Since in our model, the curvature perturbation is generated and amplified during the waterfall transition, non-Gaussian effects could potentially alter the outcomes \cite{Kawasaki:2015ppx}. However, exploring this aspect on the influence on PBH production and GW amplitude is beyond the current scope of our study and will be the focus of future publications. With the planned global network of GW detectors worldwide we will soon enter an gravitational wave
astronomy. Soon we will be able to achieve measurement precisions that are orders of magnitude better with respect to the current GW detectors. This will hopefully our the dream of testing fundamental BSM microphysics like probing the Peccei Quinn scale, a reality.

\section*{Acknowledgment}
Authors thank to Adeela Afzal and Nadir Ijaz for helpful discussions.

\bibliography{bib}
\bibliographystyle{JHEP}

\end{document}